\documentclass[preprint,11pt]{elsarticle}

\usepackage{amssymb}

\usepackage{subcaption}
\usepackage{xcolor}
\usepackage{graphicx}
\usepackage{mathtools}
\usepackage{amsmath}
\usepackage{graphicx}      
\usepackage{lineno}
\usepackage[
  colorlinks=true,
  linkcolor=blue,
  citecolor=blue,
  urlcolor=blue
]{hyperref}
\usepackage[margin=1in]{geometry}

\usepackage{mathtools}

\DeclarePairedDelimiter\floor{\lfloor}{\rfloor}

\begin{document}

\begin{frontmatter}



\title{Numerical and data-driven modeling of spall failure in polycrystalline ductile materials}

\author[inst1]{Indrashish Saha}
\author[inst1]{Lori Graham-Brady\corref{cor1}}

\cortext[cor1]{Corresponding author: \href{mailto:lori@jhu.edu}{lori@jhu.edu}}

\affiliation[inst1]{
  organization={Department of Civil and Systems Engineering, Johns Hopkins University},
  addressline={3400 N. Charles Street},
  city={Baltimore},
  state={MD},
  country={US}
}


\begin{abstract}
Developing materials with tailored mechanical performance requires iteration over a large number of proposed designs. When considering dynamic fracture, experiments at every iteration are usually infeasible. While high-fidelity, physics-based simulations can potentially reduce experimental efforts, they remain computationally expensive. As a faster alternative, key dynamic properties can be predicted directly from microstructural images using deep-learning surrogate models. In this work, the spallation of ductile polycrystals under plate-impact loading at strain rates of $\mathcal{O}(10^6s^{-1})$ is considered. A physics-based numerical model that couples crystal plasticity and a cohesive zone model is used to generate data for the surrogate models. Three architectures — 3D U-Net, 3D Fourier Neural Operator (FNO-3D), and U-FNO were trained on 
the particle-velocity field data from the numerical model. The generalization of the models was evaluated using microstructures with varying grain sizes and aspect ratios. U-FNO and 3D U-Net performed significantly better than FNO-3D across all datasets. Furthermore, U-FNO and 3D U-Net exhibited comparable accuracy for every metric considered in this study. However, training the U-Net requires almost half the computational effort compared to U-FNO, making it a desirable option for a surrogate model. Additionally, a small search problem is designed to show the effectiveness of the deep-learning models in an iterative framework requiring multiple evaluations of the spall strength, which showed a $200$ times acceleration. 

\end{abstract}




\end{frontmatter}


\section{Introduction}
Designing materials for extreme environments in the aerospace, automotive, and defense industries requires the understanding of their behavior at high strain-rates, temperatures and pressures. Quasi-static properties and low-strain-rate effects can be evaluated and quantified based on experiments such as uniaxial tension, compression or bending. However, for dynamic behavior, materials are subjected to extremely high strain rates of $\mathcal{O}(10^6 s^{-1})$ or even greater. Achieving such strain rates are possible by applying shock loading to the material through plate impact experiments ~\cite{meyers1994dynamic}. 
This work focuses on understanding the spall behavior of ductile polycrystals, which is achieved through plate impact experiments ~\cite{10.1093/pnasnexus/pgae148,MALLICK2021104065}. Spall can be defined as the dynamic failure of materials under extreme hydrostatic tensile stresses generated by the interaction of rarefaction waves inside the material. In ductile materials, spall failure is caused by the nucleation and growth of voids, which coalesce to produce cracks and ultimately cause failure ~\cite{antoun2006spall,meyers1994dynamic}. Experimentalists have found that spall strength is strongly influenced by microstructural features such as grain size, void volume fractions and grain boundary properties.  Wayne et al. identified preferred locations of void nucleation during spall at grain boundaries with misorientations in the range $25^\circ - 50^\circ$ ~\cite{WAYNE20101065} leading to intergranular failure in copper. Brown et al. found that spall occurred at grain boundaries with almost $90 \%$ probability, which reduced as failure transitioned to a mix of intergranular and intragranular upon using heat treatment and full recrystallization ~\cite{brown2015correlations}. \\
\indent Numerically, molecular dynamics (MD) has been used to study spall failure at the sub-micron scale, ~\cite{flanagan2022role, REMINGTON2018313}, as it quantifies the actual stresses associated with void nucleation and growth. However, not only are MD simulations computationally intensive, they are typically conducted at very high strain rates ($10^8-10^{11} s^{-1}$), which are orders of magnitude higher than the experimental time scales. Continuum models can be used to capture more realistic time scales. At the continuum scale, finite element studies have been conducted for isotropic ~\cite{VERSINO2018395,BRONKHORST2021102903} as well as anisotropic ductile materials ~\cite{moore2018modeling,LIEBERMAN2016270}. Krishnan et al. developed a modified Gurson-Tvergaard-Needleman based damage model to predict spall failure in copper at relatively low strain rates ~\cite{krishnan2015three}. Nguyen et al. developed a multi-scale model for dislocation dynamics-based void growth in ductile single crystal copper under dynamic loading ~\cite{NGUYEN20171,NGUYEN2021104284}. In another work, Nguyen et al. proposed a homogenized framework for ductile damage based on void growth in single crystals ~\cite{NGUYEN2020103875}. While all these works used void growth based damage models for spall failure,  Clayton ~\cite{CLAYTON20054613} and Vogler et al. ~\cite{VOGLER2008297} had used crystal plasticity (CP) and cohesive zone elements to simulate spall in extruded Tungsten alloy.  \\
\indent Numerical models grounded in physical laws are capable of delivering accurate results, but their computational cost can be prohibitive and accumulates when embedded in iterative routines such as optimization loops requiring multiple evaluations. In such cases, deep learning (DL) or machine learning (ML) models can be useful,  producing results in a fraction of the time required by the numerical models ~\cite{liu2015predictive,https://doi.org/10.1002/adma.201904845,https://doi.org/10.1002/adma.202305254,YOUSEFPOUR2025117698}. In recent years, DL and ML approaches have become popular choice for surrogate models in various applications in mechanics, specifically to reduce computational cost ~\cite{fuhg2024review}. The ML/DL models used in physical systems are usually of three types (i) physics-constrained (ii) fully data-driven (iii) a combination of data-driven and physics-constrained. Physical constraints are usually implemented using a physics-based loss function following the approach by Raissi et al. for physics-informed neural networks ~\cite{RAISSI2019686}. However, these approaches require knowledge of the exact physics of the system, which is not always readily available for complicated applications in mechanics like spallation. In these situations, fully data-driven approaches can offer significant advantages. In particular, convolutional neural network–based architectures have proven highly effective, tackling tasks that range from predicting scalar material properties to reconstructing full‐field distributions of mechanical variables~\cite{GAVALLAS2024117207,BHADURI2022109879, SAHA2024116816, IBRAGIMOVA2022103374}. A major development in the field of scientific ML has been the development of neural operator methods, of which the Fourier Neural Operator developed by Li et al.~\cite{li2020fourier} and DeepONet by Lu Lu et al. \cite{lu2021learning} are the most widely used architectures. Operator learning has proven to have a broad range of applications spanning various disciplines such as fluid dynamics~\cite{li2020fourier}, continuum mechanics~\cite{RASHID2023105444} and climate prediction~\cite{pathak2022fourcastnet}. There have been various upgrades to these architectures, for example, Wavelet Neural Operator (WNO)~\cite{TRIPURA2023115783}, U-Fourier neural operators (U-FNO) ~\cite{WEN2022104180} and U-Neural operator (U-NO)~\cite{rahman2022u} among others. However, data-driven methods can suffer from a lack of generalizability and might require a large amount of data. These issues can be addressed using some known constraints of the system for example, equilibrium conditions ~\cite{khorrami2024physics,oommen2025equilibrium}, minimization of potential energy ~\cite{ESHAGHI2025117785} or learning PDEs ~\cite{li2024physics,MANDL2025117586}. \\
\indent 
Although DL-based prediction of material responses of polycrystalline microstructures has been studied in the past ~\cite{khorrami2023artificial,khorrami2024physics,oommen2025equilibrium,IBRAGIMOVA2022103374}, these works were mostly limited to continuous stress/strain field predictions and did not involve any evolving discontinuity in the form of crack growth. Some studies have been conducted that treat cracks as discontinuities. Perera et al. used a graph neural network approach, to predict the crack growth and coalescence in a brittle medium~\cite{PERERA2022115021} while Sepasdar et al. used a U-Net type architecture to predict the post-failure crack path in fibre-reinforced composites~\cite{SEPASDAR2022115126}. However, these models do not incorporate the complexities of a problem like spall failure, in which the model must address fracture in polycrystalline inelastic materials under dynamic loading. The current work demonstrates the applicability of DL-based surrogate models for predicting this complex behavior. As a first step, a data set is generated through the development of a numerical model for simulating plate impact experiments on polycrystalline copper to predict spall failure, using the CP and cohesive zone model. Based on this data, three DL-based surrogate models are trained and compared for generalizability and accuracy, demonstrating the significant computational acceleration achieved by these models.  The paper is organized as follows. The details of the numerical model and its implementation are in Section~\ref{sec:Numerical model}. Section~\ref{sec:Method_ML} contains the details related to the DL model architectures and their training. Finally in Section~\ref{sec:results}, the numerical results related to microstructural morphology and the DL predictions are discussed followed by the application of the data-driven model in a simple search problem to demonstrate its effectiveness in an iterative framework.



\section{Methodology}
\subsection{Numerical model for the plate impact experiment}\label{sec:Numerical model}
\subsubsection{Plate impact experiments}
\begin{figure}[h!]
  \centering
         \includegraphics[width=0.7\textwidth]{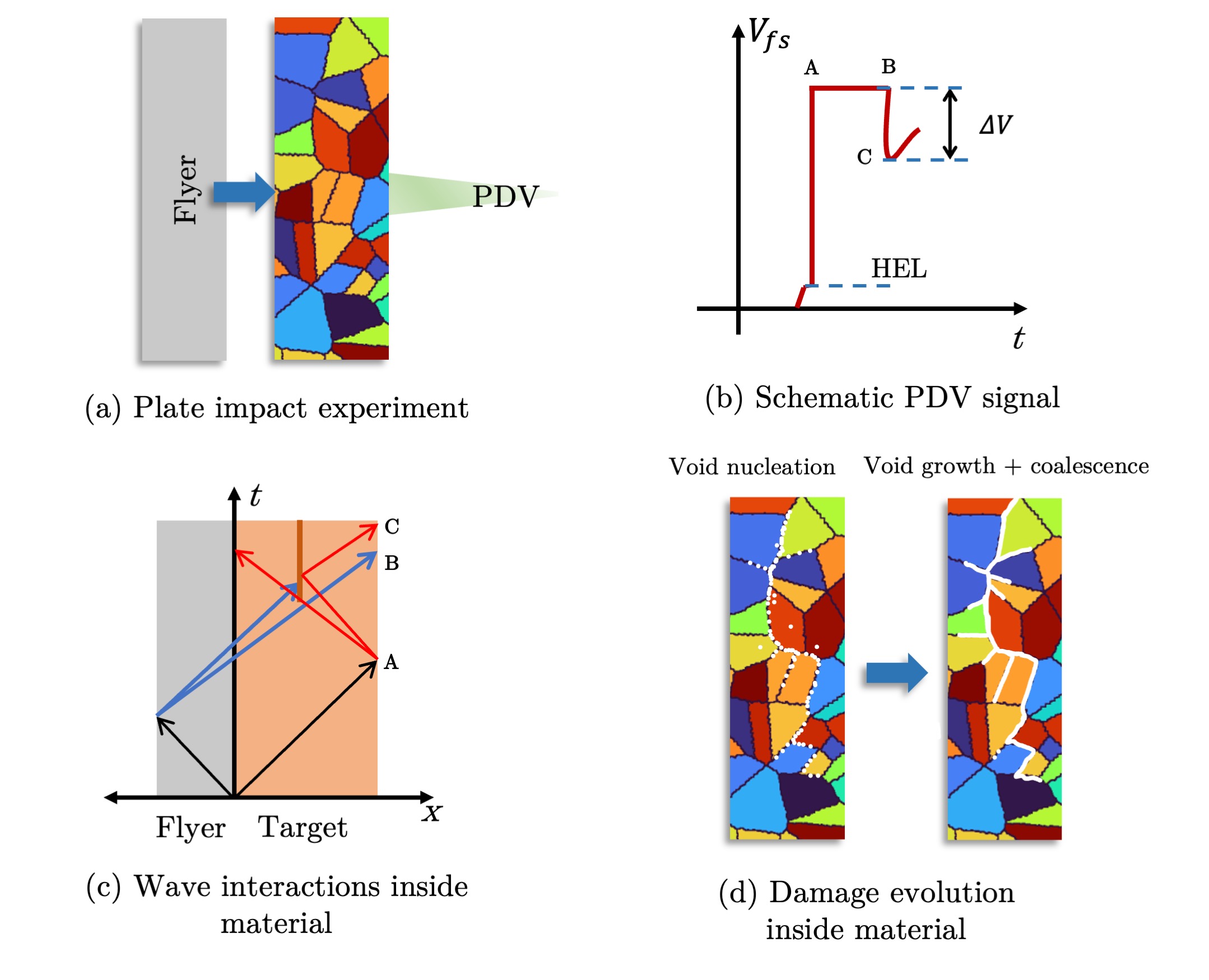}
         \caption{(a)Experimental setup depicting a flyer plate driven into a target plate made of a polycrystalline ductile material (b) A typical free surface velocity trace obtained from the back face of the target plate using PDV (c) Interaction of waves in the material and around the spall plane using a $x-t$ diagram (d) Damage growth inside the material from the nucleation of voids around grain boundaries to their coalescence and finally fracture at the grain boundaries}
         \label{fig:impact_expt}
\end{figure}

\indent Spall strength ($\sigma_{sp}$) of materials can be determined using plate impact experiments, in which a flyer plate is driven into a target plate at high velocities and the spall strength of the material is measured using Photonic Doppler velocimetry  (PDV) measurements made at the back-face of the target plates which produces the free surface velocity curves as shown in Fig.~\ref{fig:impact_expt}a and Fig.~\ref{fig:impact_expt}b ~\cite{MALLICK2021104065,WU2023144674}. The spall strength is evaluated from the free surface velocity profile using the relation,
\begin{equation}
    \sigma_{sp} = \frac{1}{2}\rho_0 C_l \Delta V 
    \label{eq:spall_eq}
\end{equation}
where $C_l$ is the longitudinal wave speed in the material, $\rho_0$ is the initial material density, and $\Delta V$ is the difference between the peak velocity and the pullback velocity shown in Fig.~\ref{fig:impact_expt}b.\\
\indent Due to the impact, compressive waves are generated in the flyer and the target plates. These waves are reflected off the free surfaces of the plates as shown in Fig.~\ref{fig:impact_expt}c as rarefaction fans. The interaction of these rarefaction fans creates large hydrostatic tensile stress. The interaction occurs inside the target material if the flyer plate is thinner than the target plate. The hydrostatic tensile stress can cause the growth of existing voids and nucleation of new ones, followed by their coalescence, leading to spall fracture as shown in Fig.~\ref{fig:impact_expt}d. 

\subsubsection{Crystal plasticity model}
\indent Since the spall fracture in a ductile material is driven by the material plasticity, crystal plasticity was used for modeling the anisotropic plastic behavior of the ductile material in the target plates. The target plate microstructures were assumed to be two dimensional Voronoi tesselations ~\cite{Voronoi} of size $200 \mu m \times 200 \mu m$. 
\begin{figure}[h!]
  \centering
         \includegraphics[width=0.6\textwidth]{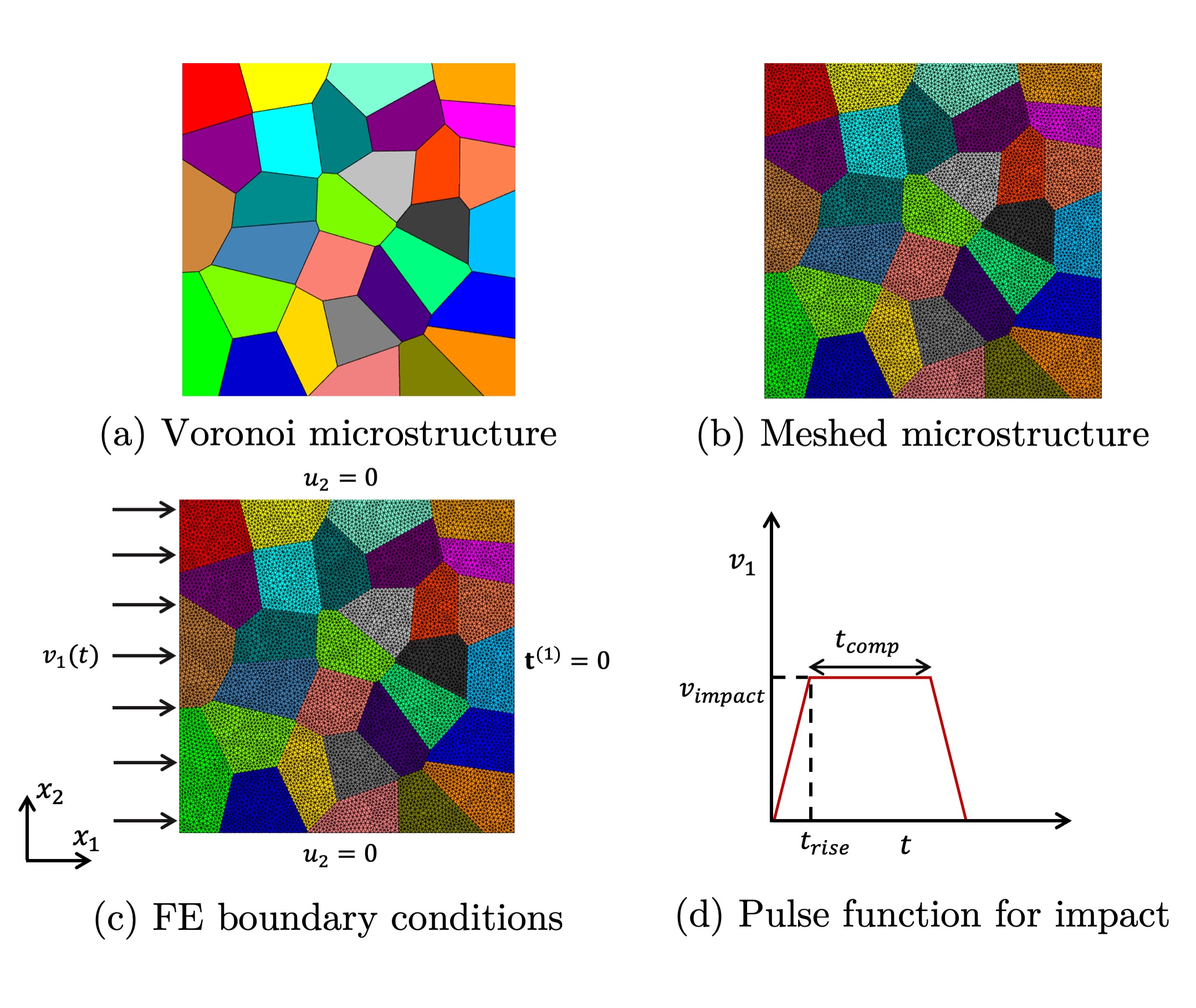}
         \caption{(a) 2D synthetic microstructure of size $200 \mu m \times 200\mu m$ and 30 grains generated using Neper ~\cite{QUEY20111729} (b) Meshed microstructure consisting of $16996$ CPE3 elements and $784$ COH2D4 elements (c) Meshed microstructure with the applied boundary conditions (d) Pulse function used to simulate impact loading. The actual function is defined in Eq.~\ref{eq:pulse_function}. }
         \label{fig:neper_micro}
\end{figure}
These Voronoi tesselation-based microstructures are generated using Neper~\cite{QUEY20111729} as shown in Fig.~\ref{fig:neper_micro}. In this work a 2D approximation along the shock direction similar to the model in ~\cite{CLAYTON20054613,VOGLER2008297} was assumed. The current work also assumes that the spall failure occurs only along the grain boundaries based on experimental observations in ductile materials ~\cite{REMINGTON2018313,brown2015correlations}. Wilkerson and Ramesh showed the possible types of failures (transgranular or inter-granular failure) based on the strain-rates and grain size of the microstructure~\cite{PhysRevLett.117.215503}. The boundary conditions and the microstructural morphologies assumed indicate that intergranular spall failure is a reasonable assumption. 
To model the grain boundary failure, zero-thickness cohesive elements were inserted at the grain boundaries in the polycrystalline microstructures.
\begin{figure}[ht]
  \centering
         \includegraphics[width=0.7\textwidth]{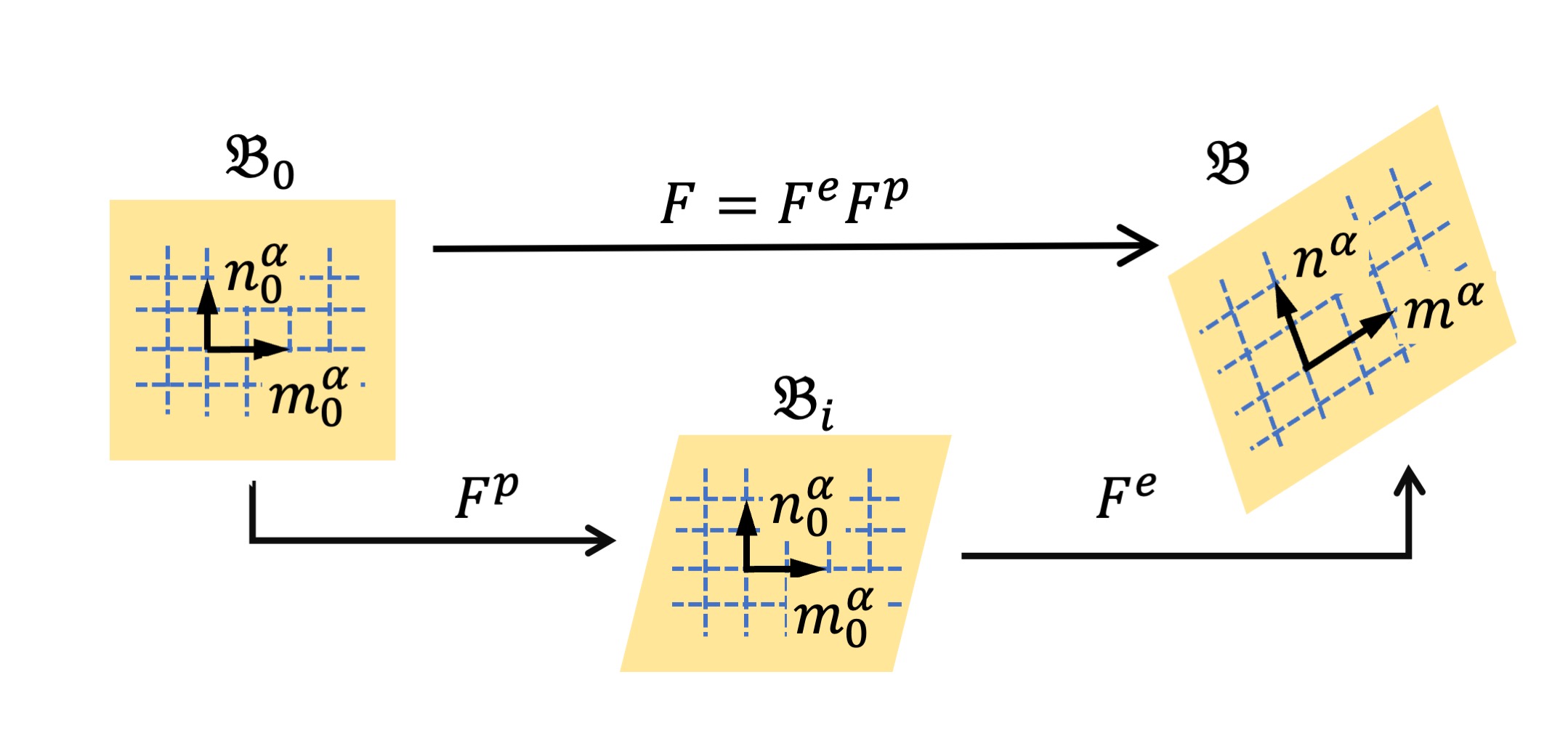}
         \caption{Decomposition of the deformation gradient following Eq.~\ref{eq:decomposition} in finite strain elasto-plasticity theory.}
         \label{fig:mul_split}
\end{figure}

\indent Following the finite strain theory of plasticity, the deformation gradient, $\mathbf{F}$ is multiplicatively decomposed into elastic and plastic parts denoted by $\mathbf{F^e}$ and $\mathbf{F^p}$, respectively:
\begin{align}
    \mathbf{F} = \mathbf{F^{e}F^{p}}
    \label{eq:decomposition}
\end{align}
$\mathbf{F^p}$ corresponds to plastic flow and is isochoric, i.e. det($\mathbf{F^p}) = 1$, and $\mathbf{F^e}$ corresponds to the elastic stretching and rigid body rotations. $\mathbf{F^p}$ transforms the reference configuration ($\mathfrak{B}_0$) to an intermediate configuration ($\mathfrak{B}_i$) and $\mathbf{F^e}$ maps it to current configuration ($\mathfrak{B}$) as demonstrated in Fig.~\ref{fig:mul_split}.
$\mathbf{F^{p}}$ can be calculated as,
\begin{equation}
   \mathbf{ \dot F^{p} = L^p  F^p }
\end{equation}
where $\mathbf{L^p}$ is defined in crystal plasticity theory as ~\cite{PEIRCE19831951,KALIDINDI1992537},
\begin{equation}
    \mathbf{L^{p} = \sum}_{\alpha = 1}^{N_s} \mathbf{\dot \gamma^{\alpha} m_0^{\alpha}\otimes n_0^{\alpha}}
\end{equation}
where $N_s$ is the number of slip systems, $\mathbf{m_0}^{\alpha}$ and $\mathbf{n_0}^{\alpha}$ are the $\alpha^{th}$ slip direction and the normal to the slip plane in $\mathfrak{B}_0$, respectively, and $\dot \gamma_0$ is calculated using a power-law formulation:
\begin{equation}
    \dot \gamma^\alpha =\dot \gamma_0 \left | \frac{\tau^\alpha-\chi^\alpha}{s^\alpha} \right |^{1/m}sgn(\tau^\alpha-\chi^\alpha)
    \label{eq:slip_law}
\end{equation}
where $\dot \gamma_0$ is the reference shear rate, $\chi^{\alpha}$ is the back-stress, $s^{\alpha}$ is the slip resistance and $m$ is the hardening exponent. The evolution equations of $s^{\alpha}$ and $\chi^{\alpha}$ are defined in ~\cite{KALIDINDI1992537}. The resolved shear stress $\tau^\alpha$ is:
\begin{equation}
    \tau^{\alpha}=\mathbf{ \sigma'} : (\mathbf{m^{\alpha}\otimes n^{\alpha}}) 
\end{equation}
where $\sigma'$ is the deviatoric Cauchy stress:
\begin{equation}
   \mathbf{\sigma'} = \mathbf{\sigma}+p_{eos}\mathbf{I} 
\end{equation}
where $p_{eos}$ is the pressure calculated from the Birch-Murnaghan equation of state ~\cite{CLAYTON20131983} based on the volume change of the material,
\begin{equation}
    p_{eos}= \frac{3B_0}{2}\left [J_e^{-\frac{7}{3}}-J_e^{-\frac{5}{3}}\right] \left(1+\frac{3}{4}(B_0'-4)\left [J_e^{-\frac{2}{3}}-1\right ] \right)
\end{equation}
$B_0$ and $B_0'$ are the bulk modulus and the pressure derivative of bulk modulus of the material and $J_e = $det($\mathbf{F^e}$). In current configuration $\mathfrak{B}$ the Cauchy stress $\sigma$ is obtained as,
\begin{equation}
    \mathbf{\sigma} = \frac{\mathbf{F^e T F^{e^T}}}{J_e}
    \label{eq:Push forward}
\end{equation}
In order to split the hydrostatic and deviatoric stresses consistently, following Krishnan et al.~\cite{krishnan2015three}, $\mathbf{F^e}$ is further split into,
\begin{equation}
    \mathbf{F^e} = \mathbf{F^e_{vol}F^e_{dev}}
\end{equation}
where $\mathbf{F^e_{dev}}= J^{-\frac{1}{3}}_e \mathbf{F^e}$. $\mathbf{F^e_{dev}}$ is used to calculate the deviatoric Cauchy stress $\mathbf{\sigma'}$. 



\subsubsection{Cohesive zone modeling}
As mentioned in the previous section, the numerical model relies on cohesive zones at the grain boundaries to simulate the fracture process. Cohesive zones were modeled using a trapezoidal traction separation law and a large cohesive energy ~\cite{SCHEIDER20031943,YUAN20141} to prevent a complete brittle failure of all the grain boundaries. The traction separation law is parameterized on cohesive energy $\Gamma$ and the displacement at failure $\delta_f$, where $\Gamma$ is the area under the traction separation curve given by,
\begin{equation}
    \Gamma = \int_0^{\delta_{f}} T(\delta) d\delta
\end{equation}

\begin{figure}[htbp]
\centering
    \includegraphics[width=0.8\textwidth]{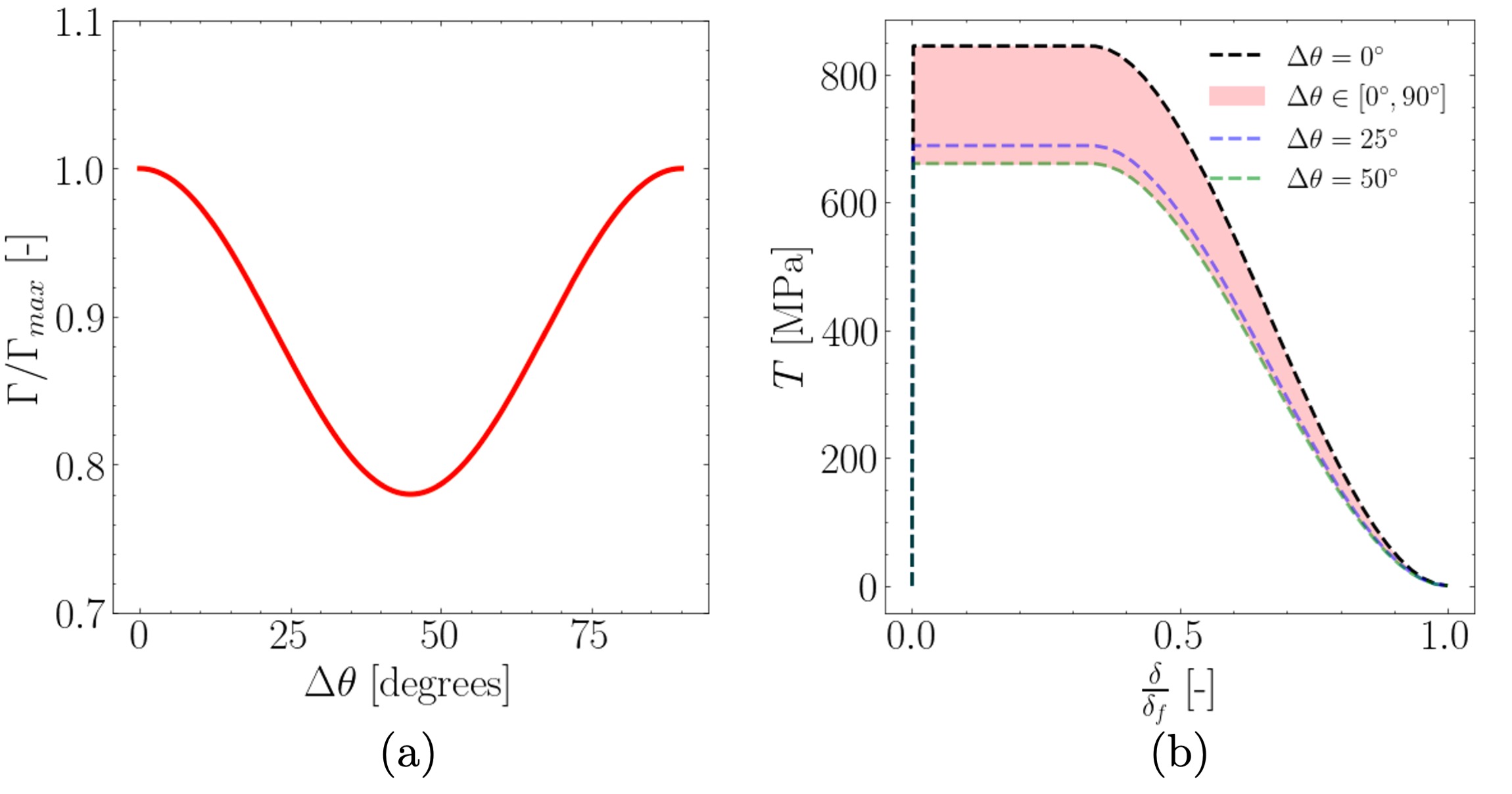}
  \caption{(a) Grain boundary cohesive energy as a function of misorientation angle between two grains. The values are normalized by $\Gamma_{max}$ (b)Traction separation law normalized on $\delta_f$ and dependent on the misorientation of grain boundaries. The pink region shows the spread due to the misorientation. The curve also shows the traction separation curve for $\Delta \theta = 0^\circ$ which is significantly higher than the grain boundaries. }
  \label{fig:TS_law}
\end{figure}

Additionally, a displacement, $\delta_0$ was fixed to mark the beginning of the plastic regime of the traction separation. $\delta_0$ is responsible for the initial stiffness of the cohesive elements. Assuming $25\% $ of $\Gamma$ is used during damage and the softening regime as a right-angled triangle, $T_{max}$ can be evaluated by~\cite{SCHEIDER20031943},
\begin{equation}
\frac{\Gamma}{4} = \frac{1}{2}T_{max}(\delta_f-\delta_1)
\label{eq: damage energy}
\end{equation}
where $\delta_{1}$ is the displacement where softening begins and is assumed to be a constant fraction of $k\delta_{f}$, with $k=\frac{1}{3}$. 
The cohesive energy can be a misorientation dependent parameter. 
Based on studies on copper by Wayne et al. ~\cite{WAYNE20101065} grain boundary failures happened for specific misorientations in copper polycrystals ($25^\circ-50^\circ$) and specifically for high energy grain boundaries.  Roy et al. had modeled grain boundary strength as a periodic function of misorientation ($\Delta \theta$) ~\cite{ROY2021104384}. Salvini et al. observed that fracture energy has a misorientation dependence which can be modeled as a convex function ~\cite{SALVINI2024103854}. Based on these observations a periodic function similar to ~\cite{ROY2021104384} was selected for the current work, which is defined as, 
\begin{equation}
\Gamma(\Delta\theta) = \frac{1}{2}(\Gamma_{max}+\Gamma_{min})+\frac{1}{2}(\Gamma_{max}-\Gamma_{min})
\cos(4\Delta\theta)
\label{eq: Coh ener}
\end{equation}
where $\Delta \theta$ is the misorientation of the grain boundaries, $\Gamma_{max}$ and $\Gamma_{min}$ are cohesive energy bounds.
Additionally, Salvini et al. had assumed $\Gamma_{min}$ is a fraction of $\Gamma_{max}$ ~\cite{SALVINI2024103854} given by,
\begin{equation}
    \Gamma_{min} = \nu \Gamma_{max}
\end{equation}
 A variation of $\Gamma$ with misorientation is presented in Fig.~\ref{fig:TS_law}a. Finally, a general functional form is selected for the trapezoidal traction separation law~\cite{SCHEIDER20031943},\\

\begin{equation}
T(\delta) =
\begin{cases}
T_{max}\frac{\delta}{\delta_0},
  & \delta \le \delta_0,\\[6pt]
T_{\max},
  & \delta_0 < \delta < \delta_1,\\[6pt]
T_{\max}\,\biggl[\,2\Bigl(\dfrac{\delta - \delta_1}{\delta_f - \delta_1}\Bigr)^3
  - 3\Bigl(\dfrac{\delta - \delta_1}{\delta_f - \delta_1}\Bigr)^2
  + 1\biggr]
  & \delta_1 \le \delta \le \delta_f.
\end{cases}
\label{eq:traction_law}
\end{equation}
 Using Eq.~\eqref{eq: damage energy},~\eqref{eq: Coh ener} and~\eqref{eq:traction_law},  a misorientation-dependent traction separation law can be obtained as illustrated in Fig.~\ref{fig:TS_law}b. $T_{max}$ values at $\Delta \theta = 25^\circ$ and $\Delta \theta = 50^\circ$ are the lowest for the range of misorientations considered which facilitates imposing the preferential failure of grains in this misorientation range, thus following the experimental observation in ~\cite{brown2015correlations}. 
\begin{table}[ht]
    \centering
    \renewcommand{\arraystretch}{1.2}
    \caption{Parameters of the traction separation law used for the plate impact simulations}
    \label{tab:myparameters}
    \begin{tabular}{c c c c c}
        \hline
        $\delta_f$ & $\delta_0$  & $\Gamma_{\max}$ & $\nu$ \\
        $[\mu m]$  & $[\mu m]$    & $[kJ/m^2]$            & $[-]$ \\
        \hline
        4.00 & $10^{-2}$ &  3.50 & 0.78 \\
        \hline
    \end{tabular}
\end{table}
All parameters related to the traction separation law have been listed in Table~\ref{tab:myparameters}.
\subsubsection{FE model implementation}
\begin{figure}[h!]
\centering
    \includegraphics[width=0.85\textwidth]{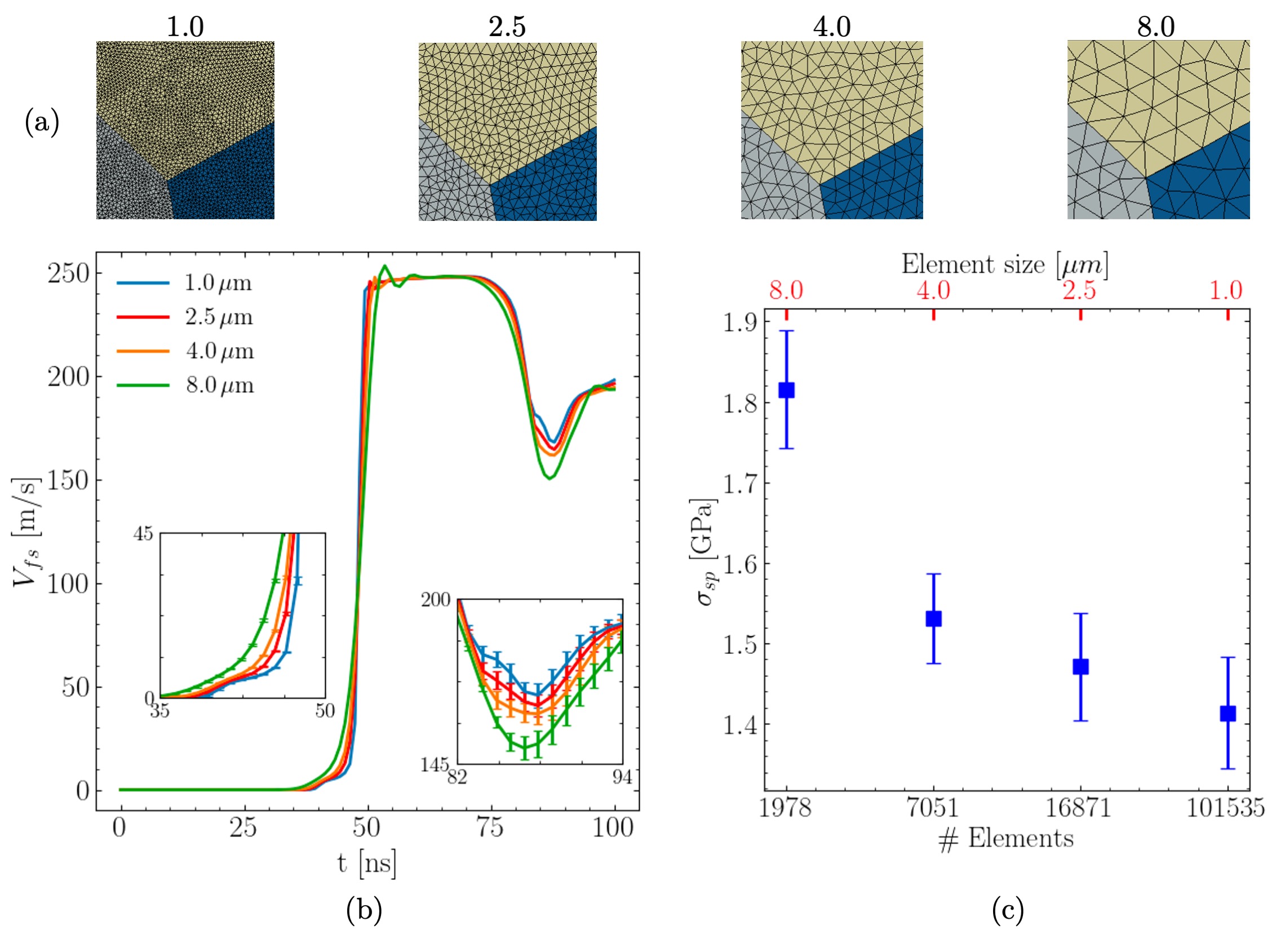}
  \caption{(a) Magnified region in a polycrystal domain showing the mesh with element sizes $1 \mu m$,  $2.5 \mu m$, $4 \mu m$ and  $8.0 \mu m$ \quad (b) Free surface velocity traces for different mesh sizes \quad (c) Spall strength variation with the change in mesh size. }
  \label{fig:mesh_converge}
\end{figure}

Microstructures imported from Neper are assigned boundary conditions using a Python script similar to the schematic present in Fig.~\ref{fig:neper_micro}c. In summary, the vertical displacement was constrained on the top and bottom edges, {\it i.e.}, $u_2(x,y = 0) = 0$ and  $u_2(x,y = L_y) = 0$,  where $L_y = 200 \mu m$. Furthermore, the free surface,  {\it i.e.}, $x = L_x$ where $L_x = 200 \mu m$ was kept traction-free. An impact velocity loading was applied at the other side {\it i.e.}, $x=0$, using a trapezoidal pulse as shown in Fig.~\ref{fig:neper_micro}d. The trapezoidal pulse can be defined as a piecewise linear function in time,
\begin{equation}
v_1(t) \;=\; V_{\mathrm{impact}}
\begin{cases}
0.5\,\dfrac{t}{t_{\mathrm{rise}}}, 
  & t \,\le\, t_{\mathrm{rise}}, \\[6pt]
0.5, 
  & t_{\mathrm{rise}} < t < t_{\mathrm{rise}} + t_{\mathrm{comp}}, \\[6pt]
0.5\,\dfrac{\,t_{\mathrm{pulse}} - t\,}
                 {\,t_{\mathrm{pulse}} - \bigl(t_{\mathrm{comp}} + t_{\mathrm{rise}}\bigr)\,}
  & t_{\mathrm{rise}} + t_{\mathrm{comp}} \le t \le t_{\mathrm{pulse}}
\end{cases}
\label{eq:pulse_function}
\end{equation}
where $V_{impact}$, $t_{rise}$, $t_{comp}$ and $t_{pulse}$ are the impact velocity, rise time, compression period and the pulse width respectively. For these simulations $V_{impact}$ was taken as $250$ m/s, $t_{rise}$ as $1$ ns, $t_{comp}$ as $34$ ns, and $t_{pulse}$ as $36$ ns ~\cite{VOGLER2008297}. The ``$0.5$" factor was used assuming that a symmetric impact of plates i.e. plates of the same material ~\cite{VOGLER2008297,davison2008fundamentals}, take place and the particle velocity is half of the impact velocity.
Equations ~\eqref{eq:decomposition} --~\eqref{eq:slip_law} were implemented in Abaqus/Explicit using user-defined subroutine VUMAT. The microstructures were meshed using 2D plane-strain elements (CPE3) and the grain boundaries were meshed using zero-thickness cohesive elements (COH2D4). The numerical model also employs a general contact algorithm at the interfaces to prevent penetration of elements during compression. All the FE simulations were run on Intel Xeon Gold CPUs (3.00 GHz).\\
\indent Explicit integration requires small increments for stable simulation and accurate results, and the cohesive elements require one or two orders of magnitude smaller increments than the bulk elements, as recommended in ~\cite{falk2001critical}. Therefore, the effective element size selected for the microstructures was $2.5 \mu m$, which resulted in about $17000$ elements on average. The time increment used for the simulation was $10^{-12}$ s. To evaluate this mesh, a convergence study was conducted, using mesh sizes of $1\mu$m, $2.5\mu$m, $4 \mu$m, and $8 \mu$m as shown in Fig.~\ref{fig:mesh_converge}a. Five samples of each mesh size were taken to understand the average behavior. The mean free surface velocity traces were compared, and it was observed that the pullback velocities in $t\in[82,94]$, became higher as the mesh became finer, see Fig.~\ref{fig:mesh_converge}b. Additionally, the mesh size affected the rise time and the Hugoniot elastic limit (HEL) as seen in $t\in[35,50]$. The coarser $8 \mu m$ mesh fails to resolve the HEL, and the trace shows a smooth rise in Fig.~\ref{fig:mesh_converge}b. Furthermore, it was observed that spall strength reduced and reaches a constant value as the mesh size reduces, as seen in Fig.~\ref{fig:mesh_converge}c. Fig.~\ref{fig:mesh_converge}c shows an approximate $6\%$ difference in spall strength between the $1 \mu$m and $2.5 \mu$m mesh size, but the computation cost is about $4$ times more for the $1 \mu$m case. Considering these trade-offs, the $2.5 \mu$m was selected for data generation.
This mesh size yields spall-strength predictions for copper that are realistic and fall within acceptable limits. All parameters used in the crystal plasticity model have been listed in Table~\ref{tab:myparameters_cp}. 


\begin{table}[ht]
  \centering
  \tiny              
  \setlength{\tabcolsep}{10pt}
  \renewcommand{\arraystretch}{0.5}
  \caption{Parameters used in the crystal plasticity modeling of copper polycrystals. The definitions of hardening parameters and their values are in~\cite{KALIDINDI1992537}.}
  \label{tab:myparameters_cp}
  \resizebox{\linewidth}{!}{%
    \begin{tabular}{cccccccccc}
      \hline
      \hline

      \noalign{\vskip 3pt}
      $\rho_0$ & $C_l$ & $C_{11}$ & $C_{12}$ & $C_{44}$& $N_{s}$ & $m$ & $\dot\gamma_0$ & $B_0$ & $B_0'$ \\
      $[kg/m^3]$ & $[m/s]$ & [GPa] & [GPa] & [GPa] & [-]& [–] & $[s^{-1}]$ & [GPa] & [–] \\
      \noalign{\vskip 3pt}
      \hline
      \noalign{\vskip 3pt}
      8960.0 & 3933.0 & 169.0 & 122.0 & 75.3 & 12 & 0.02 & $0.001$ & 140.0 & 4.79 \\
      \noalign{\vskip 3pt}
      \hline
    \end{tabular}%
  }
\end{table}

\subsection{Deep learning models for predicting material response}\label{sec:Method_ML}
 In this work, DL models based on CNNs and neural operator architectures are used to predict material response based on its microstructure. Therefore, the DL problem can be formulated as,
\begin{equation}
   \mathcal{G} : M \to V
\end{equation}
where $M  = M(x,y,t) $, denotes the microstructure, $V  = V(x,y,t)$ represents the evolution of the field variable and $\mathcal{G}$ represents the mapping between the two. The field variable considered in the current work is the horizontal component of particle velocity field ($v_p(x,y,t)$). In this work, three deep learning models have been considered -- 3D U-Net, FNO-3D and U-FNO 3D to learn this mapping ~\cite{cciccek20163d,li2020fourier,WEN2022104180}. \textcolor{black}{Although graph-based methods are an option, using these methods through the FE approach will require the additional step of generating a conformal mesh for the grains from pixelated microstructural images (such as those obtained from electron backscatter diffraction scans), requiring filtering and smoothing of grain boundaries. To avoid this added step, this work focuses on grid-based methods.}
\subsubsection{Data configuration}
The DL models were trained on the data generated from the plate impact simulations described in Section ~\ref{sec:Numerical model}. The objective of this work is to learn the microstructure-dependent spall failure of the material under a fixed impact velocity. Therefore, these models are designed to take polycrystal microstructures as input and predict the particle velocity field maps at uniformly spaced time steps. The data comprises of 2D spatial information and a third dimension containing the temporal variation. The spatial dimensions were discretized with a uniform $128 \times 128$ grid. In the temporal dimension, $32$ frames were sampled from the $100$ frames. All models were trained on microstructures with 30 equiaxed (AR = 1.00) grains using an identical train–validation–test split. The total number of unique microstructures in the dataset was $856$ of which $56$ were reserved for testing, and the remaining $800$ were split $90:10$ into training and validation sets. The $856$ samples were generated in $4$ sets distributed over $4$ nodes, each with $214$ samples. Each simulation was parallelized on $25$ cpus on every node and took around $21$ minutes. Additionally, using data-augmentation, the training dataset size was doubled by reflecting microstructures about their horizontal edge. Therefore, the final dataset consisted of $1440$ training samples, $80$ validation samples, and $56$ test samples. In the input microstructure images, each grain was represented by a unique grain-id normalized by the number of grains. The output data was normalized by dividing each pixel value by $300$ m/s, ensuring that all results remained strictly less than $1$. All the models were trained on a single NVIDIA A100 (80GB) GPU.
\subsubsection{3D U-Net model}
U-Net type architectures are capable of regression on image-type data. Therefore, it is a useful tool while predicting full-scale distributions of parameters. It uses convolutional layers to encode the high-dimensional input features in a low-dimensional space. These low-dimensional features are then mapped back to the high-dimensional space of the output parameters from the low dimension with the decoder. Additionally, the architecture consists of skip connections between different levels of the encoder and decoder to pass the high-dimensional features which the model loses after the encoding process. Using continuous loss functions, the model can be trained to learn convolutional filters to map between input and output. This work uses a 3D U-Net model which uses 3D convolutions. The 3D U-Net architecture starts with mapping input to $64$ channels and doubles in every subsequent level. The convolutional layers contain kernels of size $5\times5 \times 5$. There are $3$ levels of encoding in the current work. It was observed that kernel size of $5$ produced better results than $3$. Additionally, group normalization ~\cite{wu2018group} was used after every convolution operation with $32$ groups and the normalization was followed by a GELU activation layer ~\cite{hendrycks2016gaussian}.

\subsubsection{Fourier neural operator}
\begin{figure}[h!]
  \centering
         \includegraphics[width=\textwidth]{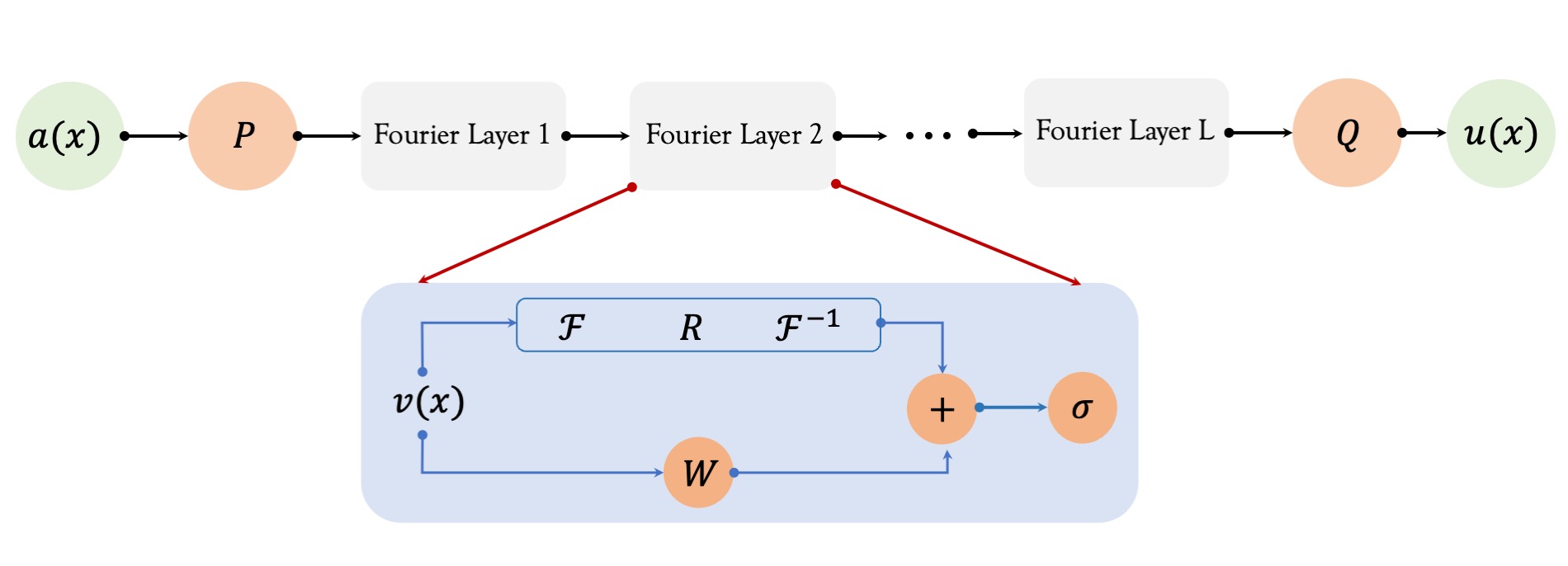}
         \caption{Fourier neural operator architecture }
         \label{fig:FNO_arch}
\end{figure}

FNO is a popular architecture used to learn mappings between functional spaces. FNO uses Fourier domain convolutions to learn long-range features. FNO consists of multiple fourier blocks, a lifting layer and a projection layer. The lifting layer maps the input to a high-dimensional space and forms input to the Fourier layers. Inside the Fourier block, the projected input is passed through two parts as shown in Fig.~\ref{fig:FNO_arch} for a one-dimensional function.
The first part goes through a Fourier transform, followed by a linear transform that transforms only lower-frequency modes and truncates out the higher-frequency modes. The low-frequency features are mapped to the physical space with the inverse Fourier transform. The other part of the input is passed through a linear transform layer that is added to the output of the inverse Fourier transform. Finally, this combination is passed through a nonlinear activation function. FNOs can learn resolution-independent features, which allows for learning in low resolution and predicting in high resolution. The original FNO paper proposed three variants FNO-1D, FNO-2D and FNO-3D. This work uses the FNO-3D version, which is a similar type of mapping as the 3D U-Net, {\it i.e.}, it does a one-shot prediction of the functions over a complete time series. This work mostly follows the FNO-3D implementation of Peng et al. ~\cite{PENG2024111063}, which is essentially based on the implementation of Li et al.~\cite{li2020fourier}. The FNO model uses $16$ independent modes and an initial lifting layer width of $40$. The current discretization of the domain with the $128 \times 128$ grid ideally requires $65$ independent modes, but to have a fair comparison with the U-FNO model described in the following section and a more efficient training, this work used $16$ modes for all three directions. The authors acknowledge that the performance could be improved with more modes, but one of the objectives of the current work is to identify a model that can produce accurate predictions at low computational cost. A study on the effect of the number of modes on FNO accuracy is presented in ~\ref{sec:appendix1}.

\subsubsection{U-FNO}
\begin{figure}[h!]
  \centering
         \includegraphics[width=\textwidth]{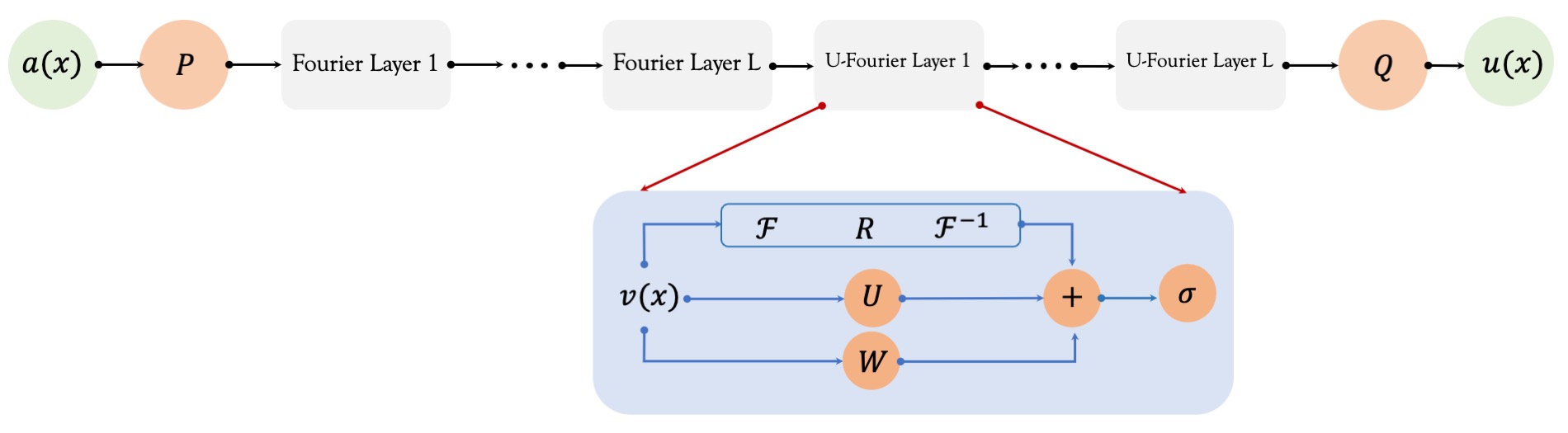}
         \caption{U-Fourier neural operator architecture with an additional U-Net path (\textit{U}) inside the U-Fourier block.  }
         \label{fig:UFNO_arch}
\end{figure}

U-FNO uses a similar architecture as the FNO. The only difference between FNO and U-FNO is the use of U-Net enhanced Fourier layers termed as the U-Fourier layers following the Fourier layers as shown in Fig.~\ref{fig:UFNO_arch}. 
The U-Fourier layers passes the input through an U-Net that helps learn the high frequency features. The introduction of CNN based U-Net prevents resolution independent learning of features which was possible in the FNO architecture. However, the authors addressed this issue in the temporal dimension by introducing down-sampling and up-sampling operations to the U-Net blocks. These operations transform the new test resolution to the original temporal resolution. A detailed description of the model can be obtained in ~\cite{WEN2022104180}. Following the recommendation in ~\cite{WEN2022104180}, the number of Fourier layers and U-Fourier layers was kept equal. U-FNO uses $16$ modes with $40$ channels lifting layer, similar to the vanilla FNO model for a fair comparison.
\subsubsection{Model training details}
The weights of the models were learnt using the Adam optimizer with a learning rate of $10^{-3}$. The relative $L_2$ loss function was used as the cost function, which can be defined as,
\begin{equation}
    \mathcal{L} =\sum_{N}\frac{ \left\| V(\mathbf{x},t)-\hat{V}(\mathbf{x},t;\theta)\right\|_2}{\left\|V(\mathbf{x},t)\right\|_2}
\end{equation}
 where  $V(\mathbf{x},t)$ is the true solution, $\hat{V}(\mathbf{x},t;\theta)$ is the model prediction parameterized by model weights $\theta$ and $\mathbf{x} \in \mathbb{R}^2$, and $N$ is the number of samples. All the models were trained for $250$ epochs with the default Adam parameters and did not require any further fine-tuning. 


\section{Results and discussion}\label{sec:results}
\subsection{Plate impact simulation}
\begin{figure}[h!]
  \centering
         \includegraphics[width=\textwidth]{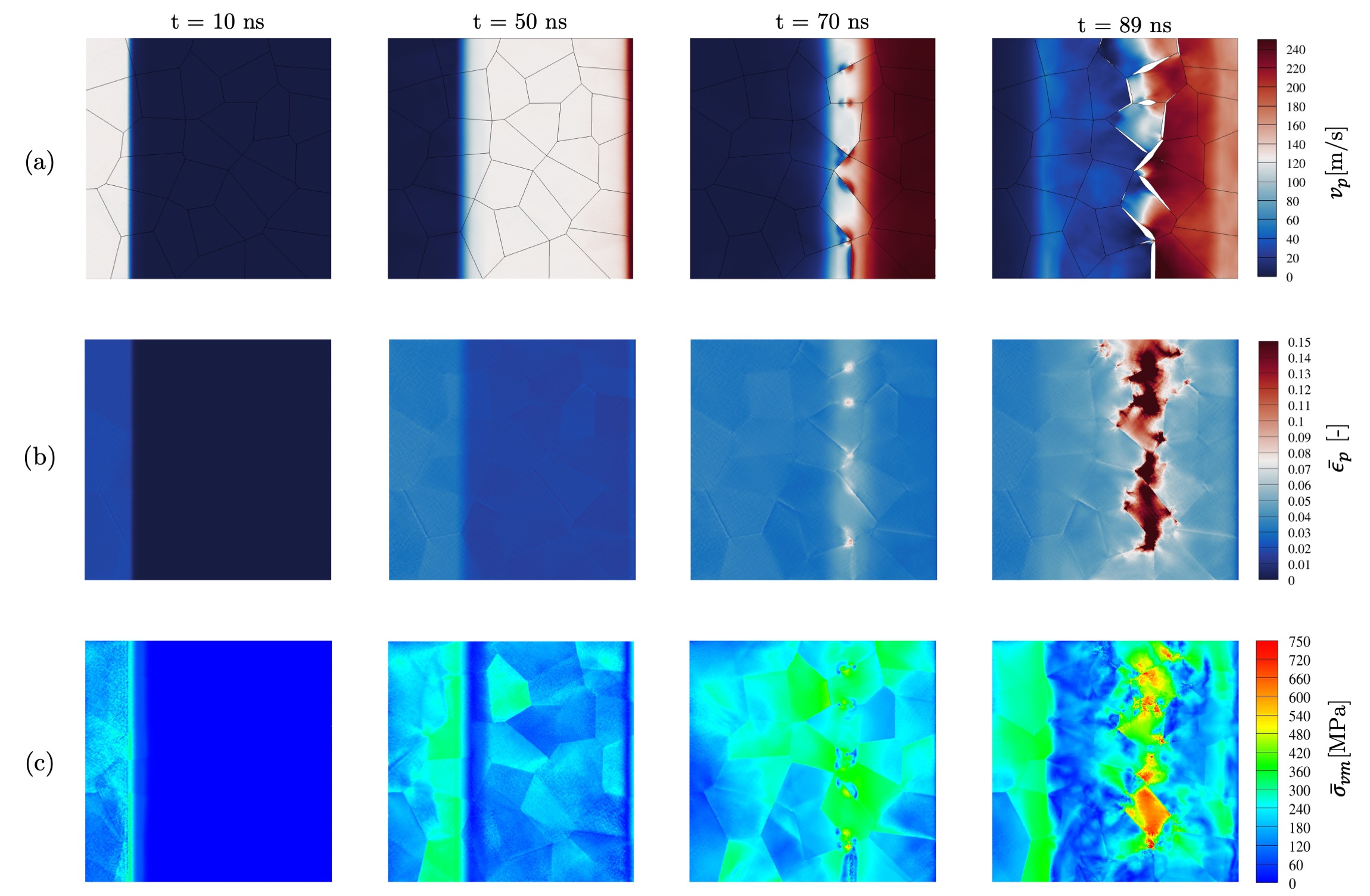}
         
        \caption{Plate impact simulation results for an impact velocity of $250$ m/s describing (a) velocity field ($v_p(x,y,t)$) (b) Equivalent plastic strain ($
    \overline{\epsilon}_p$)(c) Effective stress $(\overline{\sigma}_{vm}) $ at $t = 10, 50, 70$ and $89$ ns. These time stamps are marked on the free surface velocity traces as shown in Fig.~\ref{fig:xt_diag}c. }
         \label{fig:v1_peq}
\end{figure}

\begin{figure}[h!]
  \centering
         \includegraphics[width=0.75\textwidth]{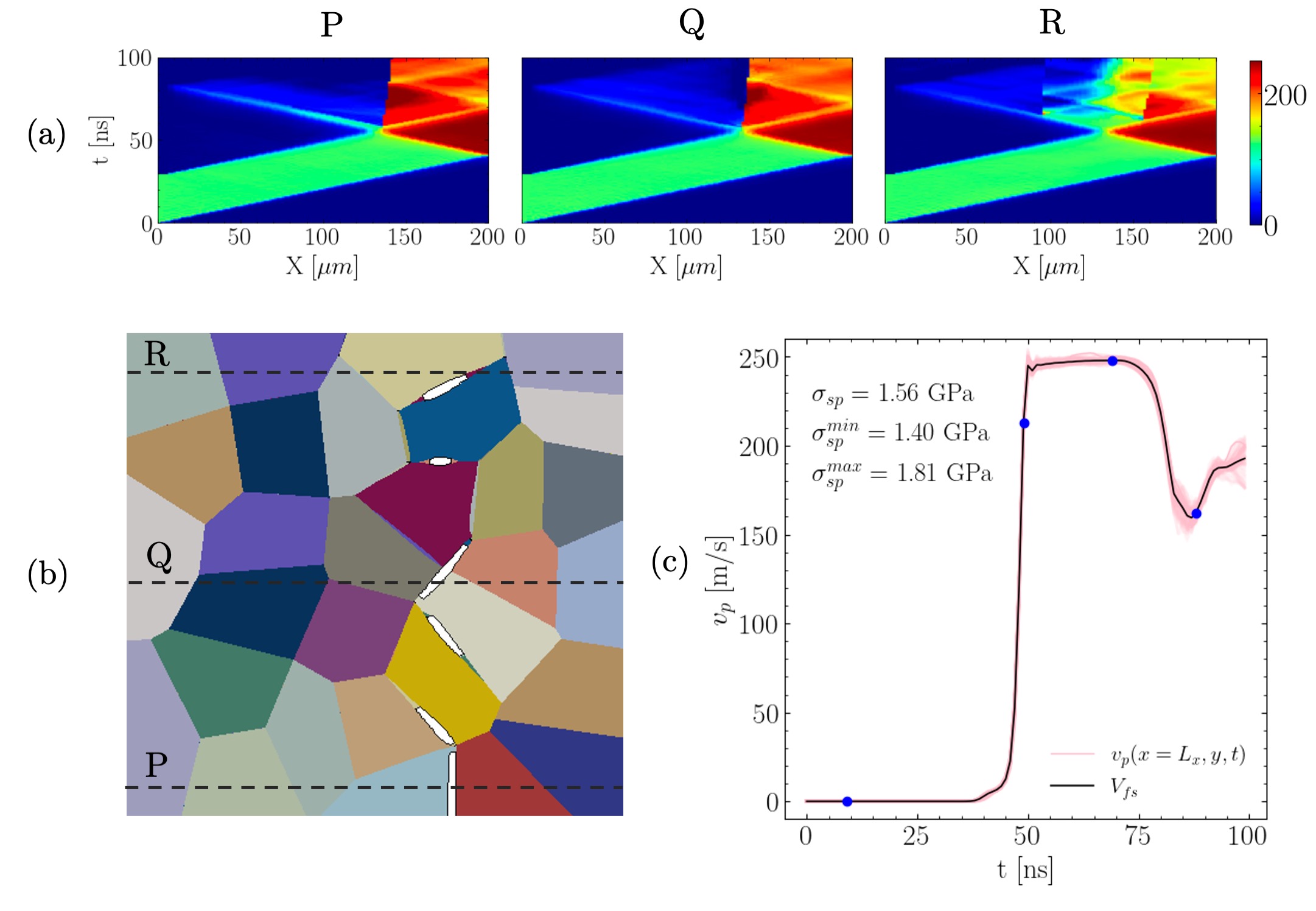}
            \caption{%
(a) Space–time ($x$–$t$) diagrams recorded at three positions (P, Q, and R) on the spalled microstructure. (b) Crack initiation and propagation in the microstructure following spall failure at $t = 100 \mathrm{ns}$. (c) Free-surface velocity traces; the shaded band denotes the range of velocities measured at different surface locations.%
}
\label{fig:xt_diag}
\end{figure}

The material response to plate impact is presented in Fig.~\ref{fig:v1_peq} at $t = 10, 50, 70, 89$ ns. The evolution of the particle velocity field in Fig.~\ref{fig:v1_peq}a describes the propagation of the compression wave, interaction of the rarefaction wave from the free surface, and finally tensile failure. After the tensile wave interaction at $t = 70$ ns, high plastic strain and effective stress concentrations occur around the spall plane, originating from the grain boundaries, as seen in Fig.~\ref{fig:v1_peq}b and c. These plastic strain concentration regions are highly correlated to the grain boundary failure zones. \\
\indent Space-time ($x-t$) diagrams are a useful method to visualize wave interactions in 1D. Three lines in the spalled microstructure were selected -- P, Q and R. The corresponding $x-t$ diagrams are given in Fig.~\ref{fig:xt_diag}. P and Q show a sharp discontinuity due to the free surface formed by the crack. But R shows wave propagation between the bulk polycrystal and the cohesive zone element, as seen in Fig.~\ref{fig:xt_diag}a. All three show similar behaviour before the interaction. 
The spalled domain at $t = 100$ ns can be visualized in Fig.~\ref{fig:xt_diag}b. It is observed that cracks have not completely separated, and only a few locations have formed free surfaces. These free surfaces are responsible for the reflection of the wave and recompression of the free surface. This recompression shows up as a pullback signal in the free surface velocity signal. The spalled domain reveals two favourable conditions for the failure: (a) orientation of grain boundaries relative to the shock direction and (b) grain boundary strength defined by $T_{max}$ in Eq.~\ref{eq: damage energy}. As stated earlier, the spall strength
$\sigma_{sp}$ is calculated based on the free surface velocity history using Eq.~\ref{eq:spall_eq}. Although experimentally, free surface velocity is measured using PDV over a small area, in this work it is calculated as ~\cite{CLAYTON20054613,VOGLER2008297},
\begin{equation}
    V_{fs}(t) = \frac{1}{L_y}\int_{y= 0}^{y = L_y} v_p(x = L_x,y,t)dy
\end{equation}
where $v_p$ is the particle velocity field in the microstructure. Using this relation, a free surface velocity history is obtained as shown in Fig.~\ref{fig:xt_diag}c. $\sigma_{sp}$ is predicted to be $\sigma_{sp} = 1.56$ GPa, which is in the range of experimental values reported in literature. Numerical simulations also show that the location of measurement of spall can affect the values. For example, if single-point values are selected at random to predict $\sigma_{sp}$, 
there is a difference of $0.41$ GPa between the minimum and maximum values. 
\subsubsection{Effect of grain morphology on spall strength}
\begin{figure}[h!]
  \centering
         \includegraphics[width = 0.8\textwidth]{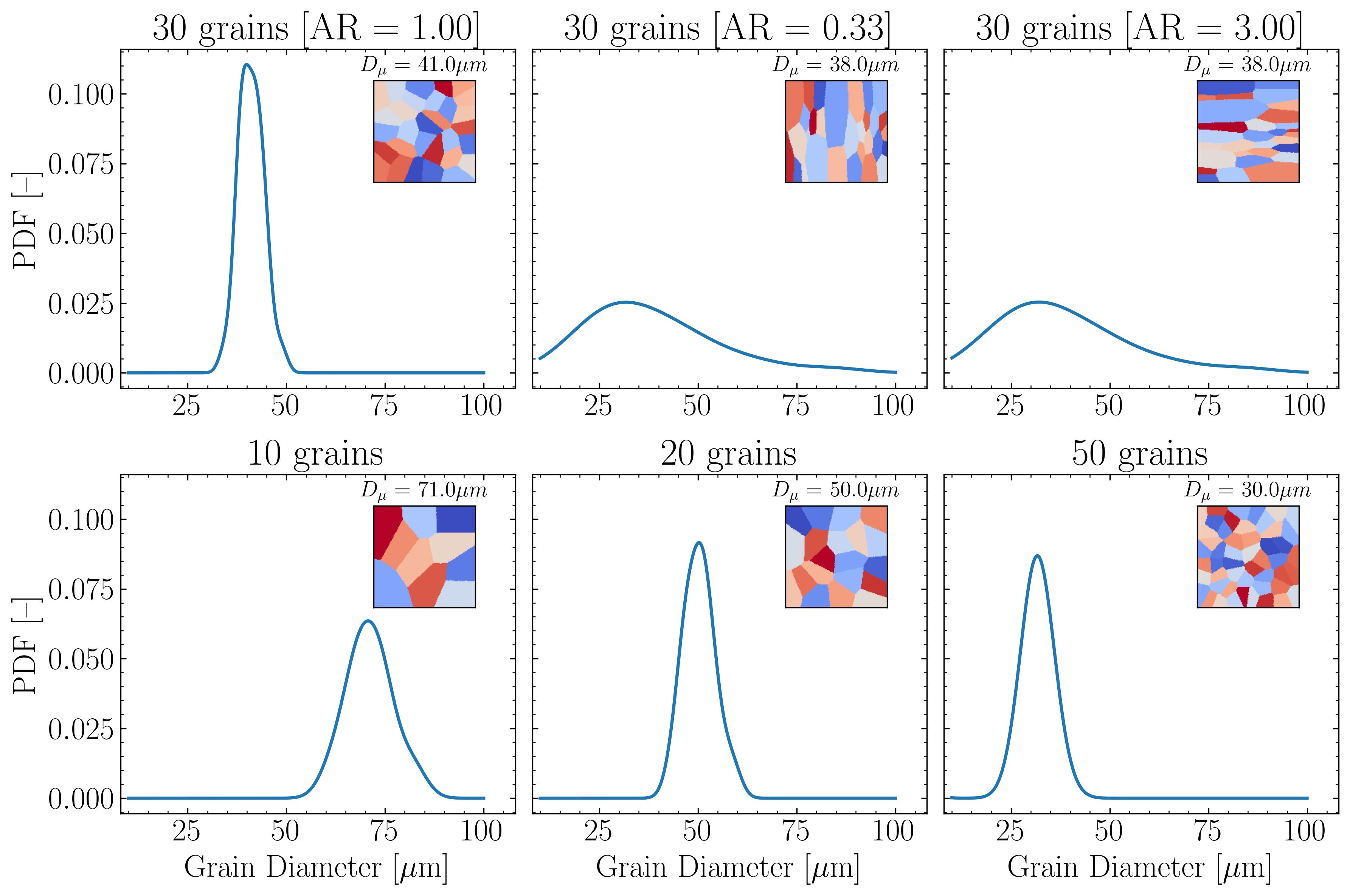}
         \caption{Grain size distribution of the different cases of microstructure with (i) 30 equiaxed grains [Training distribution] (ii) 30 grains with AR = 0.33 (iii) 30 grains with AR = 3.00  (iv) 10 grains (v) 20 grains and (vi) 50 grains.  }
         \label{fig:grain_dist}
\end{figure}

 To check for the effect of grain size,  microstructures with $50$, $30$, $20$ and $10$ grains were tested. The grain size distribution and corresponding average grain size ($D_{\mu}$) are presented in Fig.~\ref{fig:grain_dist}. The grain sizes were estimated by finding the equivalent circular diameter that fits the area of each grain. The average grain size increases from $30\mu m$ to $71\mu m$.

\begin{figure}[h!]
  \centering
         \includegraphics[width=0.85\textwidth]{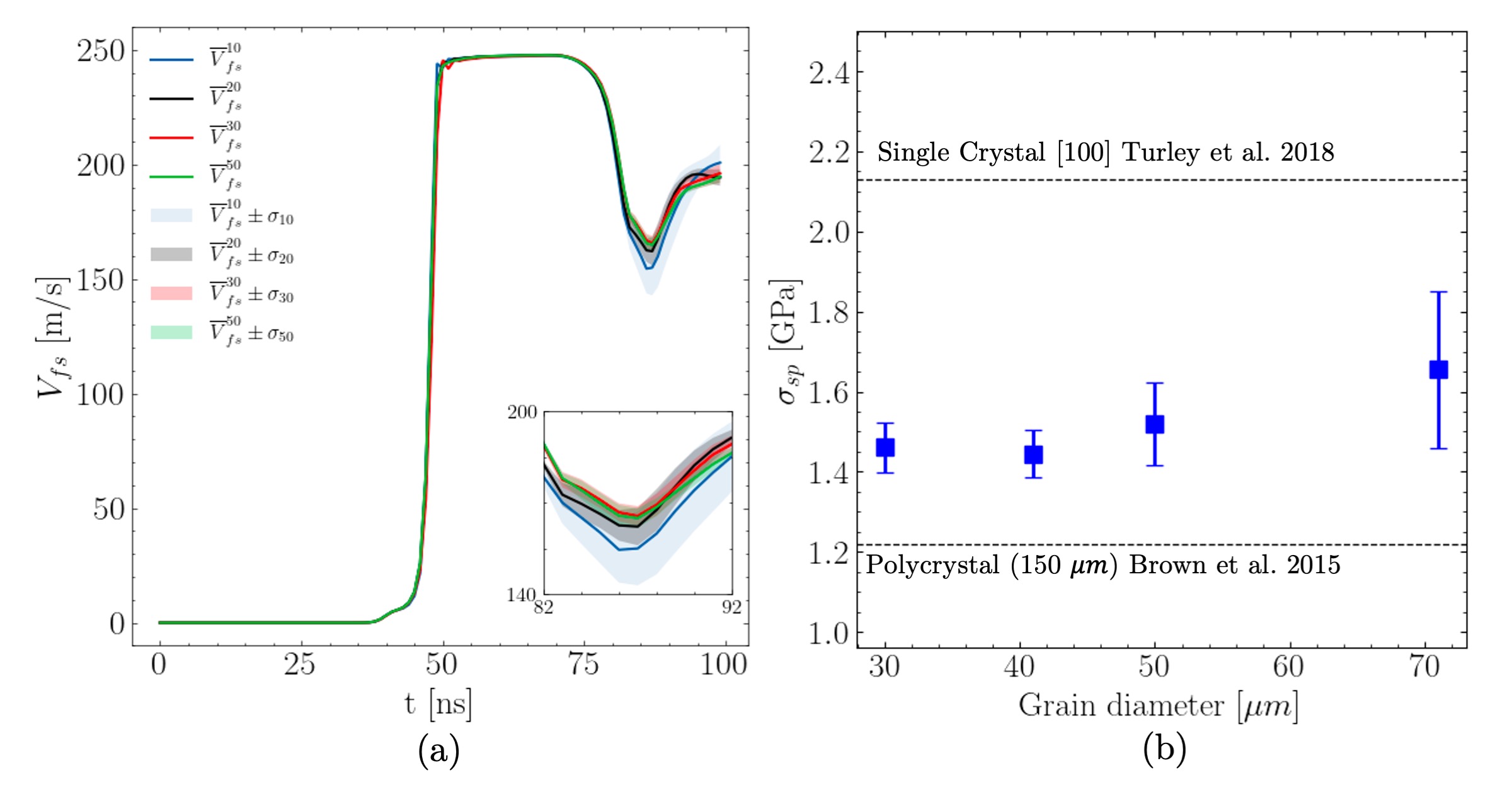}
            \caption{(a) Mean free surface velocity for different grain sizes for 10 samples and their corresponding standard deviations. In $V_{fs}^{(.)}$ and $\sigma_{(.)}$ $(.)$ represents the number of grains in the microstructure. (b) Variation of $\sigma_{sp}$ with grain size compared with data from Turley et al. and Brown et al.~\cite{brown2015correlations,Turley}.}
         \label{fig:sp_grain}
\end{figure}
The average free surface profiles and corresponding standard deviations from 10 samples are shown in Fig.~\ref{fig:sp_grain}a. The $\sigma_{sp}$ calculated from the free surface velocity profiles are plotted as a function of grain size in Fig.~\ref{fig:sp_grain}b. As expected, these values fall below the single-crystal spall strength of 
$2.13$ GPa 
measured from experiments that were conducted at $\mathcal{O}(10^6 s^{-1})$ ~\cite{Turley}, similar to the rates assumed in the FE model. The lower reference was set to $1.22$ GPa, based on ~\cite{brown2015correlations}, which corresponds to measurements on a polycrystalline material, based on lower strain rates and larger grain size than those assumed in this work. 
\begin{figure}[h]
  \centering
         \includegraphics[width=0.7\textwidth]{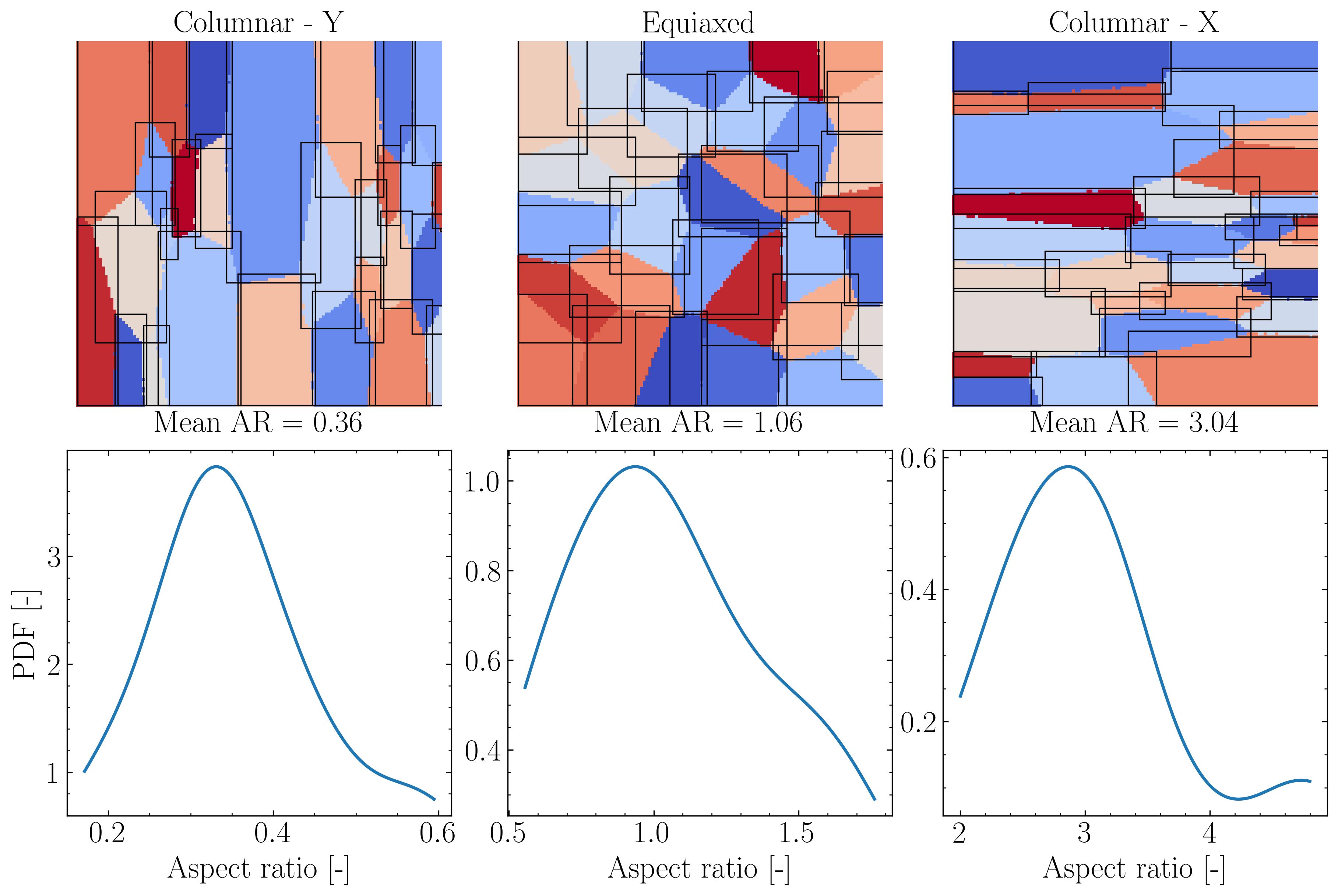}
            \caption{Microstructure grains enclosed in bounding boxes to evaluate the grain aspect ratio distribution in the microstructure}
         \label{fig:aspect_ratio_grain}
\end{figure}
Fig.~\ref{fig:sp_grain}a does not show a significant change in $\sigma_{sp}$ as the grain sizes are changed from $30 \mu m$ to $40 \mu m$. But for $D_{\mu} = 50 \mu m$ and $D_{\mu} = 71 \mu m$ there is a noticeable increase in $\sigma_{sp}$. These results suggest that there is an increase in 
$\sigma_{sp}$ as the grain size starts to approach the dimensions of the sample. For the grain sizes and strain rates considered here, this trend agrees with the observation by Wilkerson and Ramesh ~\cite{PhysRevLett.117.215503}. Furthermore, an increase in scatter of $\sigma_{sp}$ can be observed, which is closely related to the nature of the spall plane. Since there are a large number of grain boundaries close to the average vertical spall plane in cases with sizes $30 \mu m$ and $40 \mu m$, 
as seen in Fig.~\ref{fig:grain_dist}, the distance travelled by the reflected wave does not vary much. Whereas, for a grain size of $71 \mu m$, there are very few grain boundaries aligned with the vertical direction. This leads to large variability in the distance travelled by the compression waves, causing a large variation in the pullback signals, which in turn reflects in the scatter in the $\sigma_{sp}$ values.\\
\begin{figure}[h!]
  \centering
         \includegraphics[width=0.85\textwidth]{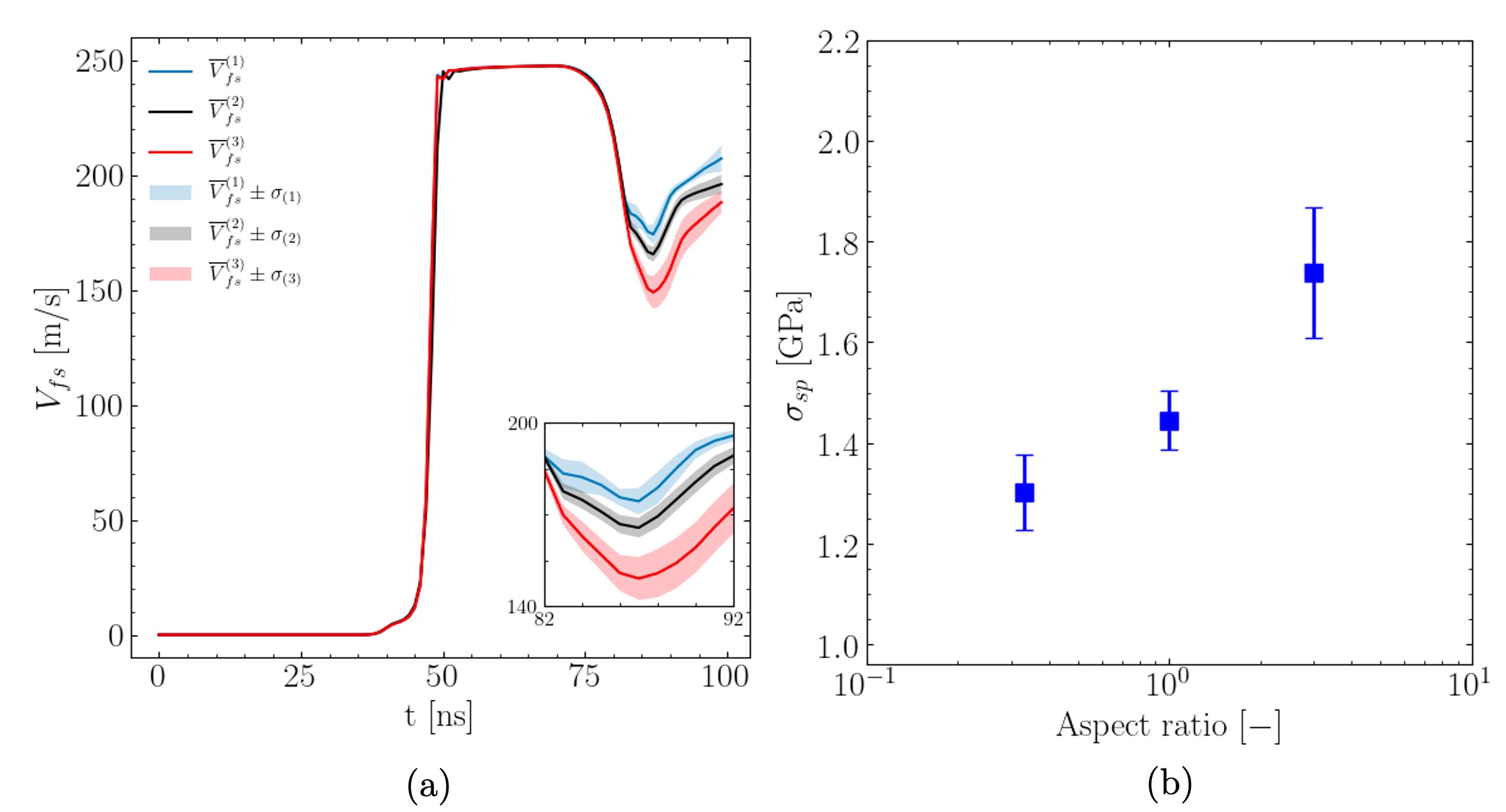}
            \caption{(a) Mean free surface velocity for different aspect ratios for 20 samples and their corresponding standard deviations. In $V_{fs}^{(.)}$ and $\sigma_{(.)}$,  $(.)$ indicates mean aspect ratio of the microstructure --(1) $ 0.33$, (2) $1.00$ (3) $3$ (b) Variation of $\sigma_{sp}$ with aspect ratio }
         \label{fig:sp_aspect}
\end{figure}
\indent The effect of grain aspect ratio was evaluated for three cases, with mean aspect ratios $0.33$, $1.00$ and $3.00$. The number of grains for the three cases was $30$. The aspect ratios were calculated based on a bounding box method, where the aspect ratio is the width-to-height ratio of the bounding box as shown in Fig.~\ref{fig:aspect_ratio_grain}. 
In Fig.~\ref{fig:sp_aspect}, a mean free surface velocity profile for $20$ samples from each case are presented. A clear distinction is observed -- microstructure with aspect ratio $0.33$ shows the highest pullback as the grain boundaries are oriented normal to shock direction, as seen in Fig.~\ref{fig:aspect_ratio_grain}, thus creating a favourable failure condition. In contrast, microstructures with an aspect ratio of $3.04$ show the lowest pullback as the boundaries are parallel to the shock direction. Fig.~\ref{fig:sp_aspect}b shows $\sigma_{sp}$ as a function of aspect ratio and shows an increasing trend. These observations show a significant effect of the orientation of grain boundaries with respect to the shock direction.

\subsection{Accelerating particle velocity predictions with deep learning}

\begin{figure}[h!]
  \centering
         \includegraphics[width=\textwidth]{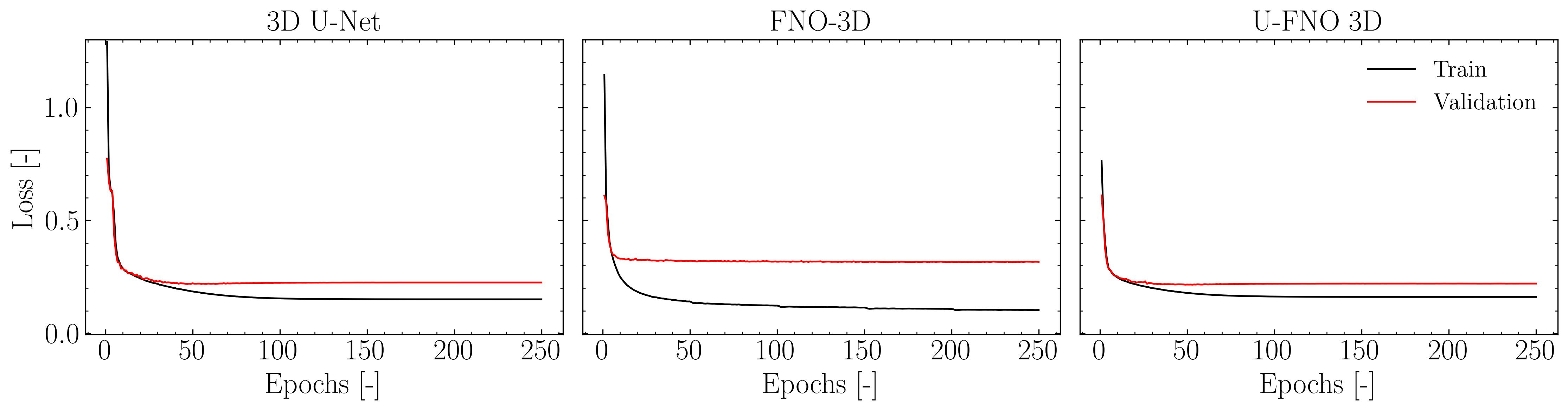}
         \caption{Variation in loss over epochs for the three DL models}
         \label{fig:loss_epoch}
\end{figure}

As mentioned in Section~\ref{sec:Method_ML}, three DL based models were used to predict particle velocity maps over the entire time history to predict the spall failure as well as estimate $\sigma_{sp}$. Spallation is a dynamic fracture event and can give rise to discontinuities in the form of grain boundary failure in polycrystal microstructure. Therefore, the models must infer the complex mechanisms of crack initiation and propagation under shock loading by leveraging explicit information from microstructure images and implicit parameters such as constant impact velocity and material properties. The three models used in this study vary greatly in their complexity, memory requirements, and computational effort required for training and evaluation. The number of model parameters, memory requirement, and computational effort have been tabulated in Table~\ref{tab:comp_param}.  All the values reported in Table~\ref{tab:comp_param} are based on the architectures selected in Section~\ref{sec:Method_ML}. The table reveals that 3D U-Net has the least number of parameters and U-FNO has the maximum number of parameters, directly affecting memory requirements. Additionally, the U-FNO takes the longest time to train at 2049 s per epoch while FNO takes the least. The 3D U-Net converges almost at the same number of epochs as the U-FNO but takes half the time. The inference times of these models on CPU were evaluated as well and 3D U-Net was found to have the least inference time.The loss plots over $250$ epochs in Fig.~\ref{fig:loss_epoch} show that U-Net and U-FNO converges with small generalization error. However, FNO-3D shows a larger generalization error compared to the other models, which is reflected in its performance as shown in the following sections. 

\begin{table}[h]
  \caption{Comparison of the computational effort for training and evaluating the surrogate models.(Note: The training time reported here is for a particular batch size, and the inference time is for each sample; `M' stands for million)}
  \centering
  \scriptsize  
  \setlength{\tabcolsep}{3 pt}
  \renewcommand{\arraystretch}{1}
  \resizebox{0.7\textwidth}{!}{%
  \begin{tabular}{ccccc}
    \hline\hline
    Model & \# Parameters  & Training Time & Inference Time \\
     & [-]  & [s/epoch] & [s] \\
    \hline
    3D U-Net     & 98M      & $\sim 1068$ & $\sim 1.70$ \\
    FNO-3D       & 104M       & $\sim 851$  & $\sim 2.80$ \\
    U-FNO 3D     & 159M      & $\sim 2049$  &$\sim  4.70$ \\
    \hline
  \end{tabular}
  }
  \label{tab:comp_param}
\end{table}


\subsubsection{Accuracy of the DL models}
\begin{figure}[h!]
  \centering
         \includegraphics[width=0.8\textwidth]{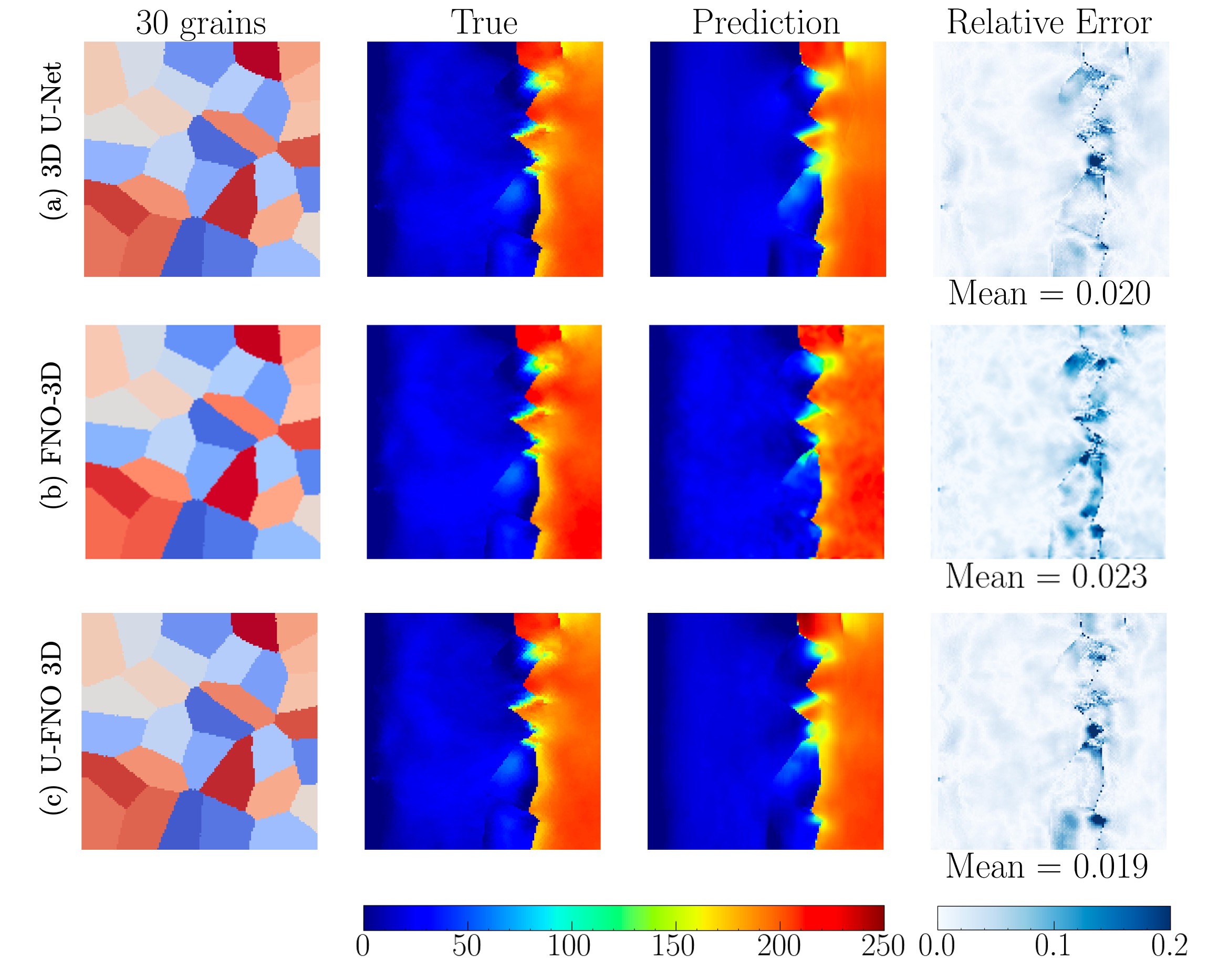}
         \caption{Velocity field predictions on unseen microstructure with 30 grains and AR = 1 at $t = 100$ ns using (a) 3D U-Net (b) FNO-3D (c) U-FNO 3D }
         \label{fig:30_grain}
\end{figure}
\begin{figure}[h!]
  \centering
         \includegraphics[width=0.5\textwidth]{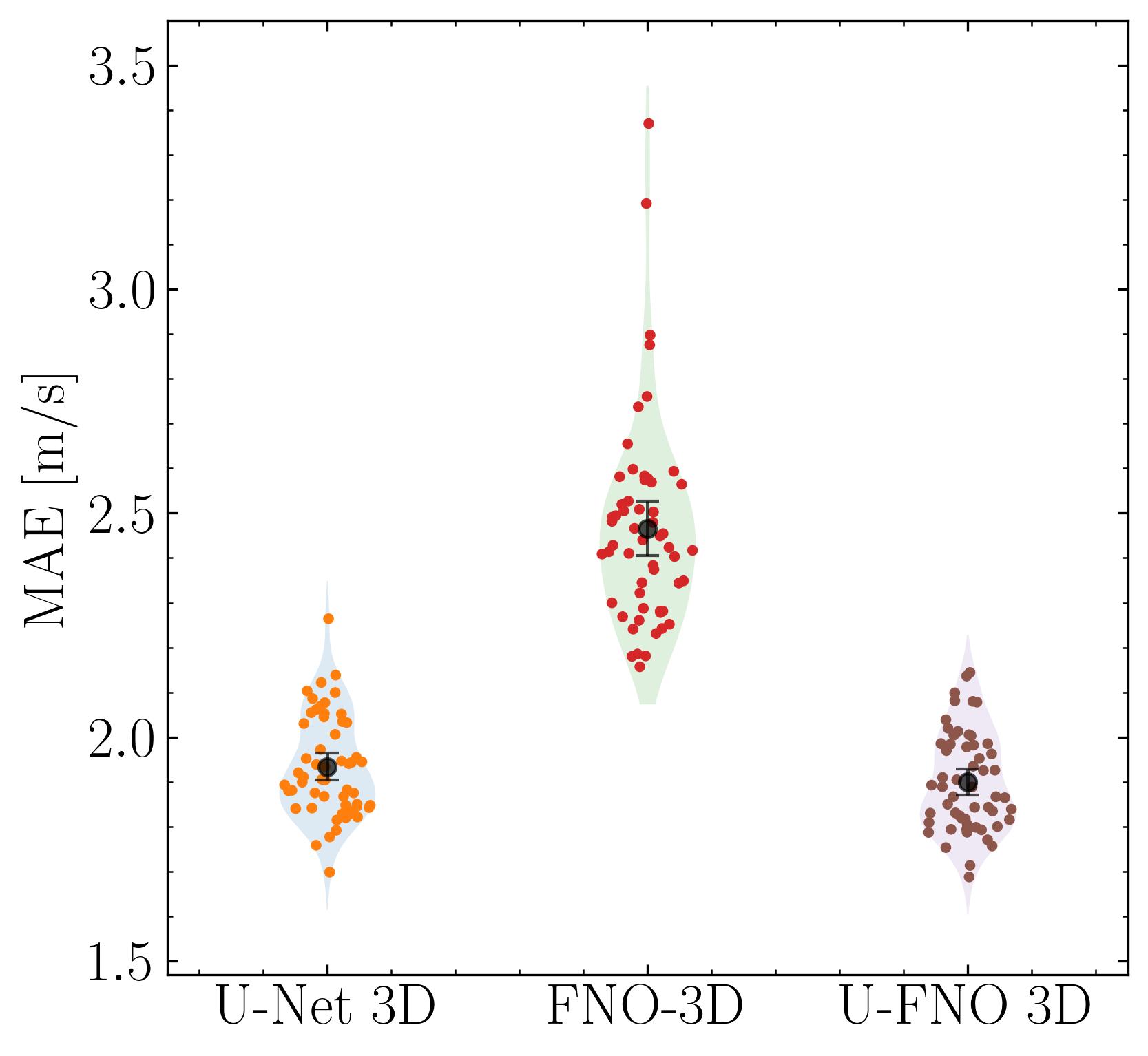}
         \caption{Mean absolute error distribution for the three models over the test dataset ($D_{\mu} = 41 \mu m$; AR $= 1$). Means are shown as black markers with $95\%$ CIs; shaded bands show kernel density estimates. }
         \label{fig:MAE}
\end{figure}
The velocity maps corresponding to microstructures with the same grain size and aspect ratio as the training data were predicted using  (i) 3D U-Net (ii) FNO and (iii) U-FNO and the results are presented in Fig.~\ref{fig:30_grain}. Since the predictions are in the form of time series of 2D maps, only the velocity map from the final step has been presented for all the models. The predicted maps are mostly similar to each other. However, the U-Net and U-FNO perform slightly better than the FNO, which has a higher local relative error. Upon close inspection of the bottom region near the spall plane, FNO fails to predict some grain boundary regions and ends up smoothening the region but these boundaries are present in the 3D U-Net and U-FNO predictions. The common characteristic of all the predicted maps is the accumulation of high error around the grain boundary region, which is primarily due to the resolution of the grids used for generating the maps. Conversion of a conformal unstructured mesh to a regular grid creates step-like grain boundaries, and if the model tries to fit these non-physical features, it can cause overfitting. To understand the absolute error levels in the predictions, the mean absolute error (MAE) for each sample in the test data has been illustrated in Fig.~\ref{fig:MAE} which shows clearly that the U-Net and U-FNO predictions are more accurate compared to FNO and show less scatter over the mean. The mean error over all the samples for U-Net and U-FNO was observed to be around $1.90$ m/s while FNO had a mean error of $2.46$ m/s.\\
\indent Previous comparative studies on U-Net type models and neural operators like FNO have shown that the operator architectures have consistently performed better than CNN architectures on polycrystal stress data~\cite{kapoor2022comparison} and solution of PDEs~\cite{li2020fourier}. One major difference in the present work is that, whereas previous datasets featured output fields that are continuous or have small gradients, this study employs discontinuous, fractured domains characterized by large gradients. Additionally, some studies predicting PDE solutions use a few initial time steps as input to forecast the next steps ~\cite{li2020fourier,rahman2022u}. In most cases, these initial steps are correlated with the predicted steps. In the current problem, the initial frames are significantly different from the later frames. To use such a training strategy, the FE model would need to run for a longer time to generate meaningful initial frames, thus nullifying the DL models' purpose. Therefore, these models have to learn a much more difficult mapping without any initial state information other than the microstructure.

\subsubsection{Effect of dataset size on model performance}
\begin{figure}[h!]
  \centering
         \includegraphics[width=\textwidth]{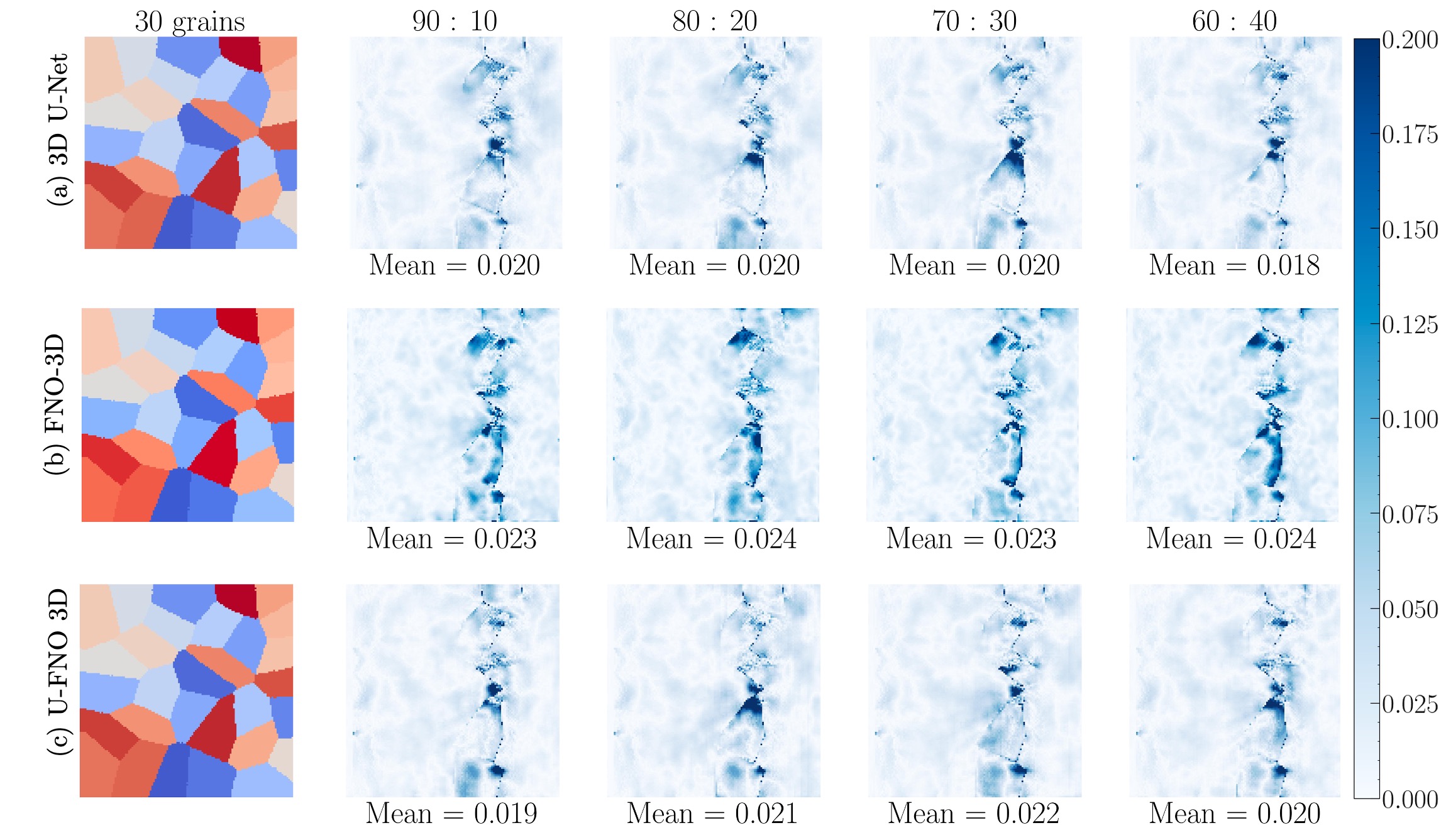}
         \caption{The relative error maps for different dataset splits (a) 3d U-Net (b) FNO-3D (c) U-FNO 3D. 3D U-Net and U-FNO 3D shows comparable performance whereas FNO predictions shows local regions of high error around grain boundaries. These results are for a 30 grain microstructure with AR =1. }
         \label{fig:30_grain_datacomp}
\end{figure}
\begin{figure}[h!]
  \centering
         \includegraphics[width=0.8\textwidth]{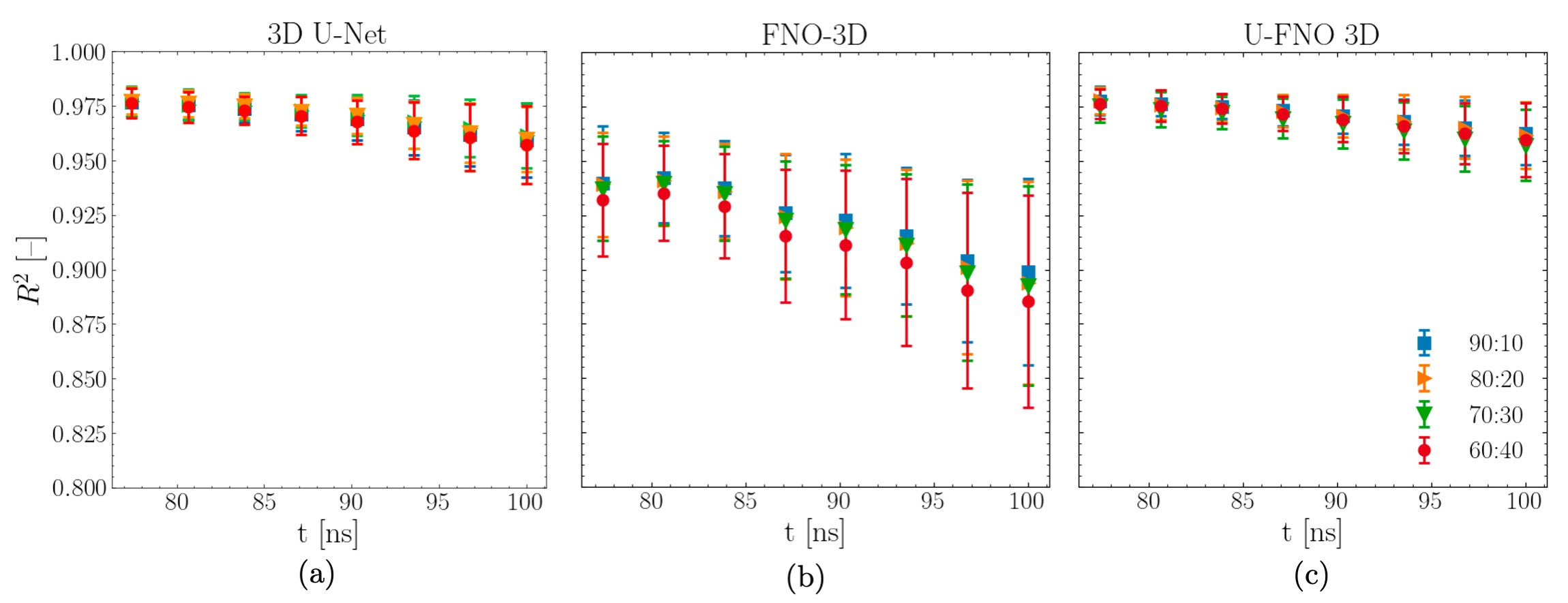}
         \caption{$R^2$ variation over frames from $t = 77$ ns to $t = 100$ ns for different train–validation splits in (a) 3D U-Net, (b) 3D FNO, and (c) 3D U-FNO.}
         \label{fig:r2_values_time}
\end{figure}

The neural operator architectures and the 3D U-Net architectures are known to be ``data-hungry". Since the numerical model is computationally expensive, generating a large amount of data can be challenging. Hence, a study on finding the optimal dataset size was conducted for four cases training-validation splits -- (i) $90:10$ (ii) $80:20$ (iii) $70:30$ and (iv) $60:40$. All the analyses were performed on the same set of test data which was separated before splitting into train and validation datasets. The variation of $R^2$ across the time series for all the test samples was evaluated, as shown in Fig.~\ref{fig:r2_values_time}. The three cases in Fig.~\ref{fig:r2_values_time}a-c show a decreasing trend of $R^2$ with time. This could be attributed to the complex emergence of cracks and their growth over time. 
Similar to earlier observations, U-FNO and 3D U-Net predictions are comparable for all cases of training dataset size. FNO produces $R^2$ values that increase somewhat as the size of the training dataset increases, but it is consistently lower than that found using the other two models. \\
\subsubsection{Testing models on microstructures with varying morphology}
To check for the generalizability limit of the models, they were tested on microstructures with morphological features different from the training distribution. In the current work, five cases were considered -- (i) 30 equiaxed grains (ii) 30 grains with AR = 0.33 (iii) 30 grains with AR = 3.00 (iv) 10 grains (v) 20 grains and (vi) 50 grains. Grain size and aspect ratio distributions presented in Fig.~\ref{fig:grain_dist} and Fig.~\ref{fig:sp_aspect} show the morphological differences between each case and the training case. \\ 
\begin{figure}[h!]
  \centering
         \includegraphics[width=0.9\textwidth]{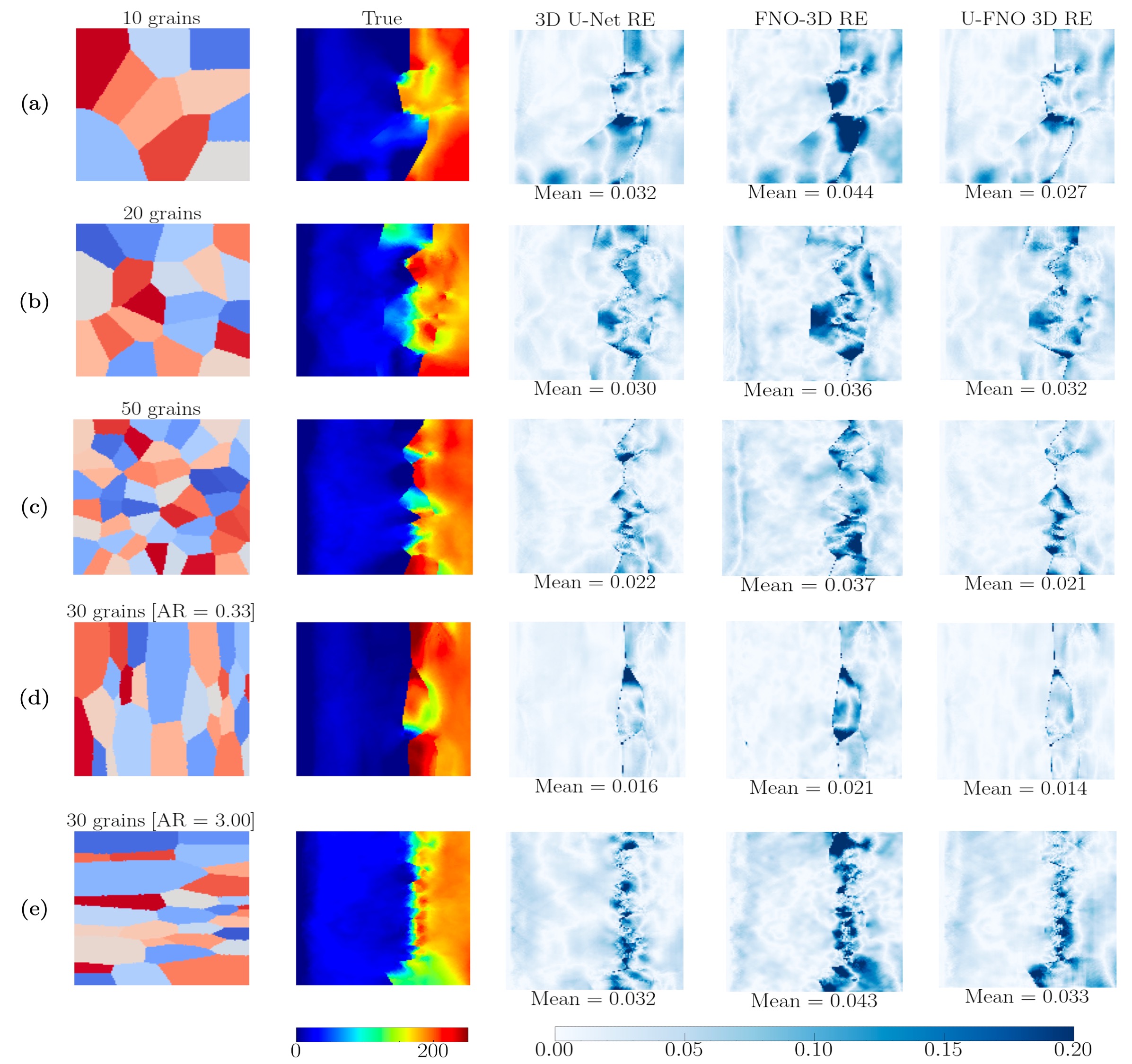}
         \caption{Comparison of model performance for different morphological variations at $t = 100$ ns in microstructure with (a) 10 grains (b) 20 grains (c) 50 grains (d) 30 grains with AR = 0.33 and (e) 30 grain with AR = 3.00. The models show better generalization to microstructures morphologically similar to training data which is expected in any deep learning model.}
         \label{fig:morpho_learn}
\end{figure}
\indent Fig.~\ref{fig:morpho_learn} shows results for generalization to different morphologies. In the first case of a microstructure with $10$ grains, the grain aspect ratios are similar to the training data, but the grain sizes are significantly larger. The relative error maps show that the models struggle to predict the regions where cracks have not fully developed, as shown in Fig.~\ref{fig:morpho_learn}a. U-FNO performs better than the other two models. This example is challenging for these models because in the regions where the waves interact, there are no grain boundaries, which is different from the cases seen while training. However, for microstructures with 20 and 50 grains, the distribution of the localized errors are similar to the training case as seen in Fig.~\ref{fig:morpho_learn}b and Fig.~\ref{fig:morpho_learn}c. This is because of the grain boundary density in the rarefaction wave interaction region is similar to the training examples.\\
\indent Fig.~\ref{fig:morpho_learn}d and Fig.~\ref{fig:morpho_learn}e illustrate predictions for two microstructures with lower and higher aspect ratio respectively, compared to the training microstructures. The microstructure with AR = 0.33 has grain boundaries aligned vertically, making it easier for the model to predict the spall plane. Hence, the models show small local errors, with U-FNO outperforming U-Net and FNO as seen in Fig.~\ref{fig:morpho_learn}d. The microstructure with  AR = 3.00 has very few grain boundaries in the wave interaction region, so the models exhibit a higher level of error compared to the lower aspect ratio.  
\begin{figure}[h!]
  \centering
         \includegraphics[width=1\textwidth]{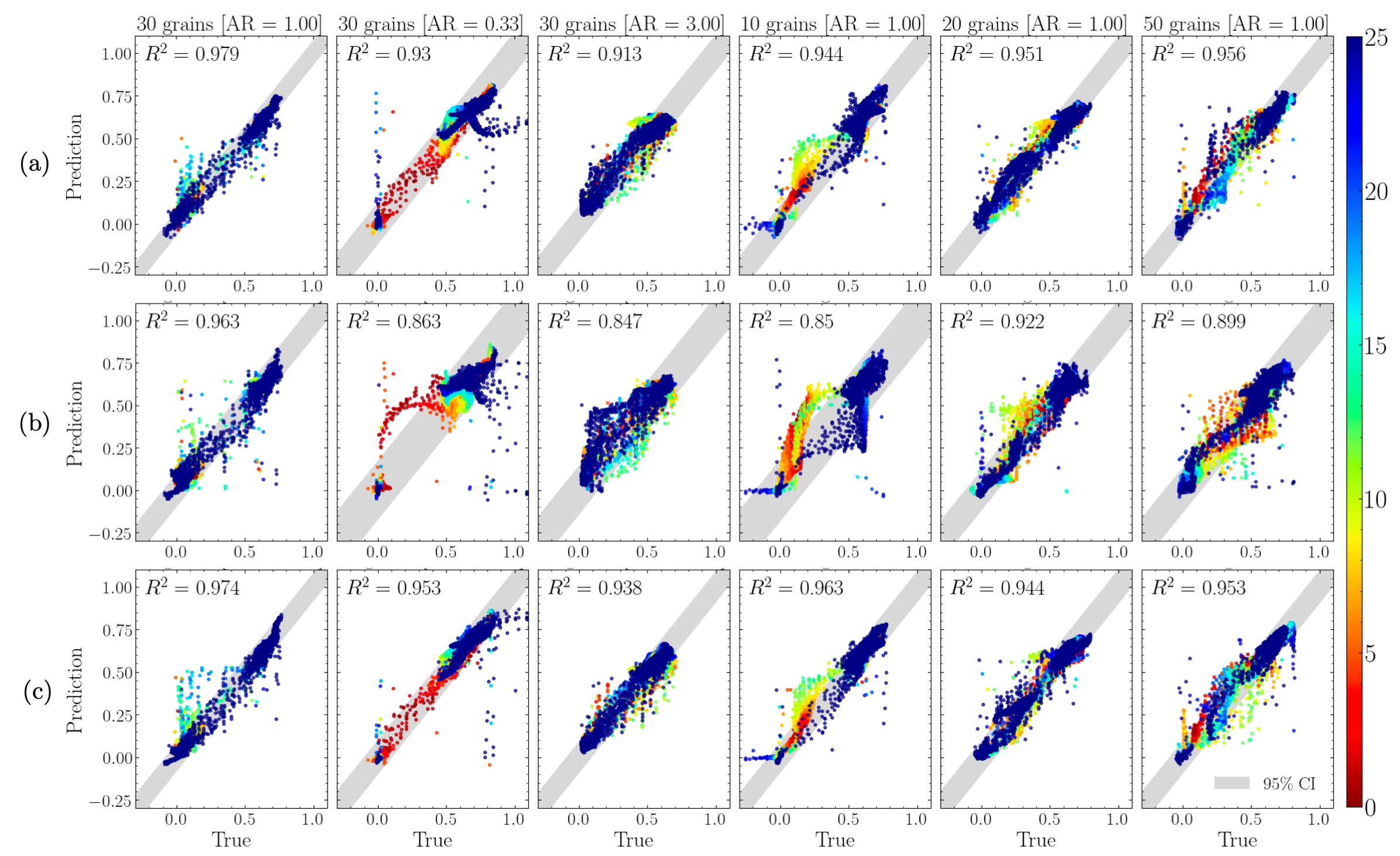}
         \caption{$R^2$ values and parity plots (between pixel values of scaled true and scaled predicted values of particle velocity) for different test microstructures in Fig.~\ref{fig:grain_dist} for (a) 3D U-Net (b) FNO 3D (c) U-FNO 3D. The points are colored based on their distance from the spall region. U-FNO 3D consistently shows better performance compared to 3D U-Net and FNO 3D.}
         \label{fig:r2_values}
\end{figure}

\indent Figure~\ref{fig:r2_values} presents parity plots of scaled particle velocity predictions and true pixel values for all test cases (from Fig.~\ref{fig:morpho_learn}) to quantify model error distributions and assess generalizability. The pixel values are from a region between the  Each plot shows a $95$ \% confidence interval: narrow bounds indicate low scatter, as seen for 3D U-Net and U-FNO 3D (Figs.~\ref{fig:r2_values} a,c), whereas FNO-3D consistently exhibits wider intervals. 
 Among the microstructures with equiaxed grains (i.e. AR = 1.00), 30, 20 and 50 grains are the best cases. Since the 30-grain case is from the training distribution, it has higher $R^2$ values and less scatter. The 20 and 50 grains cases are morphologically closer to the training distribution and thus have similar error spread and $R^2$ values greater than $0.95$. The microstructure with AR = 0.33 is an interesting case, as it shows high $R^2$ values for 3D U-Net and U-FNO close to the accuracy of training distribution but there is some heteroscedasticity in the data as seen from the vertical spread at very high and low values. This behaviour is consistent across all models. Comparison of $R^2$ values and the spread of the points shows that microstructures with 10 grains and 30 grains (AR = 3.00) prove to be the most challenging cases. These trends validate the hypothesis regarding what the model has learnt through the training.  Overall, the $R^2$ trends suggest that U-FNO and 3D U-Net outperform the FNO and show better generalization capacity to microstructures of out-of-distribution morphology.\\
\indent Although the three models show reasonable capacity to learn the full-scale particle velocity distribution, it is important to evaluate their performance in predicting spall strength, which is one of the objectives of the current work.
\begin{table}[ht]
  \caption{Spall strength comparison for 4 grain sizes. The mean and standard deviations are written following $<Predicted/True>$.(Note: (i) (*) indicates that the case is from the training distribution (ii) Fewer grains imply larger grain size as shown in Fig.~\ref{fig:grain_dist}.)}

  \centering
  \scriptsize                  
  \setlength{\tabcolsep}{4.2pt}  
  \renewcommand{\arraystretch}{1.5}
  \resizebox{\textwidth}{!}{%
    \begin{tabular}{l|ccc|ccc|ccc|ccc}
      \hline\hline
      & \multicolumn{3}{c|}{30 grains*}  
      & \multicolumn{3}{c|}{10 grains}  
      & \multicolumn{3}{c|}{20 grains}  
      & \multicolumn{3}{c}{ 50 grains}\\
      \cline{2-13}
      Model 
      & Mean & Std Dev & Error
      & Mean & Std Dev & Error
      & Mean & Std Dev & Error
      & Mean & Std Dev & Error\\
      & [GPa] & [GPa] & [\%]
      & [GPa] & [GPa] & [\%]
      & [GPa] & [GPa] & [\%]
      & [GPa] & [GPa] & [\%]\\
      \hline
      3D U-Net  
        & 1.45/1.45 & 0.05/0.09 & 0.01 & 1.45/1.63 &0.05/0.21 & 10.82 & 1.47/1.51 & 0.05/0.11 & 2.44  & 1.45/1.46 &0.02/0.06 & 0.49  \\
      FNO-3D    
        & 1.44/1.45 & 0.06/0.09 & 0.84 & 1.49/1.63 & 0.07/0.21 & 8.17 & 1.49/1.51 & 0.07/0.11 & 1.42 & 1.39/1.46  & 0.03/0.06         & 4.14    \\
      U-FNO 3D  
        & 1.44/1.45 & 0.05/0.09 & 0.88 & 1.42/1.43 & 0.05/0.21 & 12.61 & 1.43/1.51 & 0.06/0.11 & 5.40 & 1.47/1.46 & 0.03/0.06& 0.48    \\
      \hline
    \end{tabular}%
  }
  \label{tab:grain_size}
\end{table}

Spall strengths evaluated from the predicted and true data are evaluated using Eq.~\ref{eq:spall_eq} and are presented in Table~\ref{tab:grain_size} for varying grain sizes and Table~\ref{tab:aspect_ratio} for varying aspect ratios. It should be noted that subsampling in the temporal space for the purposes of discretizing the training data might change the predictions of $\sigma_{sp}$. Therefore, for consistent comparison, the ``True" results are similarly subsampled before calculating $\sigma_{sp}$ for true and predicted cases. The $\sigma_{sp}$ values are calculated from an ensemble of $10$ data points for each case of grain size and aspect ratio. 
The lowest errors are reported for the training distribution i.e. 30 grains with AR = 1 as seen in Tables~\ref{tab:grain_size} and ~\ref{tab:aspect_ratio} with U-Net having the least error.  For other 20 and 50 grain cases, the models predict $\sigma_{sp}$ reasonably with a maximum error of 5.4\% in the 20-grain case for U-FNO 3D. Although it is hard to conclude which model performs the best, it is seen that all the models struggle in the 10-grain case, as seen in the previous analyses. It is also observed that the standard deviations for the predicted data (in-distribution as well as out-of-distribution) are underpredicted by a factor of ~2 in most cases, which stems from the models learning to minimize a mean value. They perform somewhat poorly the behavior at points where the response is extreme, as can be observed in the parity plots in Fig.~\ref{fig:r2_values}.
\begin{table}[h]
  \caption{Spall strength comparison for 3 aspect ratio cases AR = 1, 0.33 and 3.00. The mean and standard deviations are written following $<Predicted/True>$. (Note: (*) indicates that the case is from the training distribution.)}
  \centering
  \scriptsize  
  \setlength{\tabcolsep}{4.2pt}
  \renewcommand{\arraystretch}{1.2}
  \resizebox{0.9\textwidth}{!}{%
  \begin{tabular}{l|ccc|ccc|ccc}
    \hline\hline
      & \multicolumn{3}{c|}{AR = $1^*$} & \multicolumn{3}{c|}{AR = 0.33} & \multicolumn{3}{c}{AR = 3.00} \\
     \cline{2-10}
Model   & Mean & Std Dev & Error  & Mean & Std Dev & Error  & Mean & Std Dev & Error   \\
      & [GPa] & [GPa] & $[\%]$ & [GPa] & [GPa] & $[\%]$ & [GPa] & [GPa] & $[\%]$ \\
    \hline

3D U-Net  & 1.45/1.45     & 0.05/0.09      & 0.01       & 1.32/1.29 & 0.04/0.08 & 1.89  &1.56/1.74 & 0.04/0.13      &10.30      \\  
    
FNO-3D  & 1.44/1.45 & 0.06/0.09      & 0.84     & 1.37/1.29     & 0.06/0.08 & 5.82 & 1.52/1.74 & 0.04/0.13      &12.31      \\  
U-FNO 3D  & 1.44/1.45 &0.05/0.09      & 0.88     & 1.34/1.29 &0.08/0.08  & 3.17  & 1.57/1.74 &  0.02/0.13     &9.75      \\  
    \hline
  \end{tabular}
  }
  \label{tab:aspect_ratio}
\end{table}
Similar observations are made for the varying aspect ratio cases in Table~\ref{tab:aspect_ratio}, which shows maximum error for AR = 3.00. However, qualitatively, the models do retain the trend of increasing $\sigma_{sp}$ with increasing AR, as was observed in the finite element results in Fig.~\ref{fig:sp_aspect}c.  

\subsubsection{Application in a small-scale search problem}
To demonstrate the utility of the DL model in iterative design problems, a small-scale search problem is set up. This simple search problem is designed in  a way that it requires multiple evaluations of the FE or the DL model to evaluate the spall strength. The search space is comprised of $90$ microstructures with three different grain sizes  and the task is to obtain the microstructure with maximum $\sigma_{sp}$. The search space for this example only includes the cases for which the U-Net is known to be accurate, and thus contains only grain sizes of $30, 41$ and $50\mu m$.
\begin{figure}[h!]
  \centering
         \includegraphics[width=0.8\textwidth]{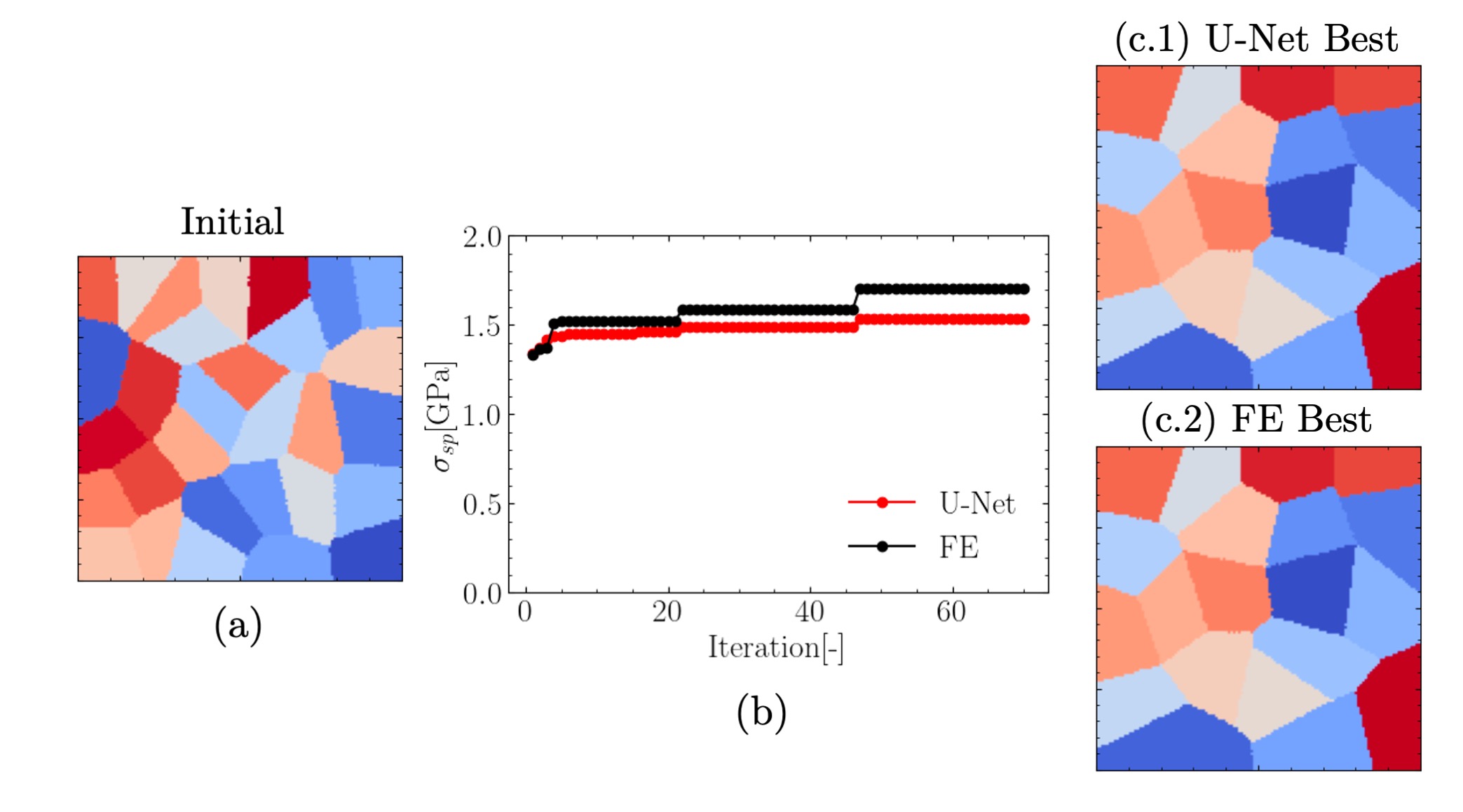}
         \caption{(a) Initial candidate microstructure chosen randomly (b) Convergence to an optimum candidate (c.1) Microstructure selected by FE based estimation (c.2) Microstructure selected by U-Net based estimation }
         \label{fig:search_example}
\end{figure}
As a well-established method for optimization in low-data search spaces, Bayesian optimization (BO) is selected here to perform a search for the optimum microstructure with maximum $\sigma_{sp}$, by implementing a Random Forest (RF) approach that parallels procedures described in ~\cite{frazier2018tutorial}. 
The algorithm is implemented using the \href{https://scikit-optimize.github.io/stable/}{\textit{scikit-optimize}} package. It is initiated by a ``warm-up" step with $10$ samples to train the RF estimator, followed by the actual optimization process in $80$ iterations. As the model moves through the iterations, the RF estimator gets trained on the evaluated cases. The training data could be either the FE or U-Net estimates of spall strength.  \\ 
BO is carried out with the two sets of data: one set from the U-Net model, and one set from the FE model. The algorithm converges in less than $70$ iterations as seen in Fig.~\ref{fig:search_example}b. 
Both the U-Net and FE-based BO reach the same optimum microstructure as seen in Fig.~\ref{fig:search_example}c. However, it should be noted that BO requires evaluation of $\sigma_{sp}$ in out-of-distribution data with the U-Net, which results in some inaccurate $\sigma_{sp}$ values. That said, the relative trend in the prediction due to morphological changes is preserved, consistent with observations in the previous section, so both methods reach the same optimum. The search example used pre-existing FE results, 
so an estimate of the time taken can be evaluated by multiplying the average time taken for each FE simulation (approximately $21$ minutes using $25$ CPUs) and the number of evaluations, which totals about $1680$ minutes. The U-Net-based BO takes around $6.12$ minutes thus achieving over $200$ times acceleration.\\
\indent Further studies with a more sophisticated optimization method over a continuous search space of grain size and aspect-ratio is required to find a truly optimum microstructure, which is beyond the scope of the current work and will be addressed in future work. This simplified example is used only to demonstrate how the U-Net model can be used to identify optimal microstructures at significantly reduced computational cost.

\section{Conclusions}
 The primary goal of this study was to develop a numerical model for spall failure in polycrystalline copper with reasonable physical assumptions and train a deep learning model to act as a surrogate of the numerical model and provide a faster evaluation of parameters. The faster evaluation of parameters is important in situations that demand multiple evaluations of a parameter, for example, in optimization loops for designing microstructures with a desired spall strength.\\
\indent The numerical model based on cohesive zone elements predicted $\sigma_{sp}$ values within experimental limits and showed some of the experimentally observed dependencies on microstructure morphology. However, some assumptions, such as using an isosceles trapezoidal pulse for impact, might not represent the exact wave interactions inside the material and affect the slope of the post-plateau region of the free surface velocity curve. A more accurate alternative is to explicitly model the flyer plate and its interaction with the target plate. However, these models require significant computational effort, making data generation for the DL models infeasible. \\
\indent
Three data-driven ML models were tested on the data generated from the spall failure simulations: (1) 3D U-Net (2) FNO-3D (3) U-FNO 3D, all of which are popular architectures used in learning dynamical systems. It was observed that the U-FNO and 3D U-Net performed better than the FNO-3D, and U-FNO was slightly better than 3D U-Net on the test dataset from the training distribution. Evaluating the test dataset (unseen data) from different distributions of grain sizes and aspect ratios, the models showed similar performance in terms of local error and $R^{2}$ values for the cases in which the morphological conditions remained similar to the training distribution. However, if the morphology of the microstructures varied significantly from the training data, as for the $10$ grain case and AR = 3.00 case, the models made less accurate predictions and showed lower $R^{2}$ values. The performance of the U-FNO and 3D U-Net was comparable in terms of every performance metric considered. However, looking beyond error metrics, an important aspect of these models is the computational cost and memory overhead involved in training. It was observed that training the U-FNO was significantly more expensive compared to the U-Net and FNO. The computational cost is reflected in the training time per epoch, which is almost $2.5$ times more than that of FNO. Although 3D U-Net and U-FNO perform significantly better than FNO-3D, it should be mentioned that the number of modes used in FNO was less than ideal and could explain the low accuracy. However, using more modes would require more computational effort and make the training computationally expensive.      \\ 
\indent As mentioned previously, these DL-based surrogate models have an important application in an iterative microstructure design framework. Design loops fully based on FE models are highly inefficient, and computation costs can rise significantly with the complexity of the models. A hybrid DL-FE based optimization could be an efficient alternative to the fully FE based optimization. To that end, a small-scale search problem was designed to show the effectiveness of the DL models. It was observed that the search task could be completed about $200$ times faster than the FE based method using the DL model. However, a full-scale optimization might have more complications since it will be working over a continuous search space, unlike the example selected here, and the acceleration might be smaller based on that. \\
\indent The DL models have an uncertainty associated with the predictions due to their ``black-box" nature. If the models can capture the correct relative trends with the variation of the input parameters, these models can reduce the initial search space of candidate microstructure and accelerate the overall process. The FE model can serve as a regularizer for the optimization by validating a small subset of the DL predicted candidates.


\section{Acknowledgements}
``AI-Driven Integrated and Automated Materials Design for Extreme Environments (AMDEE)" project sponsored by the Army Research Laboratory under Cooperative Agreement Number W911NF-23-2-0062. The views and conclusions contained in this document are those of the authors and should not be interpreted as representing the official policies, either expressed or implied, of the Army Research Laboratory or the U.S. Government. The U.S. Government is authorized to reproduce and distribute reprints for Government purposes notwithstanding any copyright notation herein. The authors gratefully acknowledge Jacob Diamond, Dr. Piyush Wanchoo and Dr. Ashwini Gupta at Hopkins Extreme Materials Institute (HEMI) for the many discussions and invaluable insights that shaped these results.\\
\indent This work was carried out at the Advanced Research Computing at Hopkins (ARCH) core facility  (rockfish.jhu.edu), which is supported by the National Science Foundation (NSF) grant number OAC1920103.
\section{Data availability}

All data and codes required for training the DL models will be available on GitHub at \href{https://github.com/Indrashish95/Learning-spall-failure-with-Neural-Operators}{Learning-spall-failure-with-Neural-Operators}.

\appendix
\setcounter{figure}{0}
\renewcommand{\thefigure}{A.\arabic{figure}} 
\section{Effect of number of Fourier modes on FNO accuracy}
\label{sec:appendix1}
The FNO results presented in the manuscript use $16$ modes in every dimension which might affect the accuracy of the model. Since the dimension of the data is $128\times128\times32$, a maximum of $ \floor*{\frac{N}{2}}+1$ modes in each dimension can be used. To understand the effect of higher number of modes, the FNO-3D model was trained with $16, 24$ and $40$ spatial modes keeping the number of temporal modes fixed to $16$. This study could not be extended to $65$ modes since the GPU ran into memory issues for the chosen batch size.
\begin{figure}[h!]
  \centering
         \includegraphics[width=0.5\textwidth]{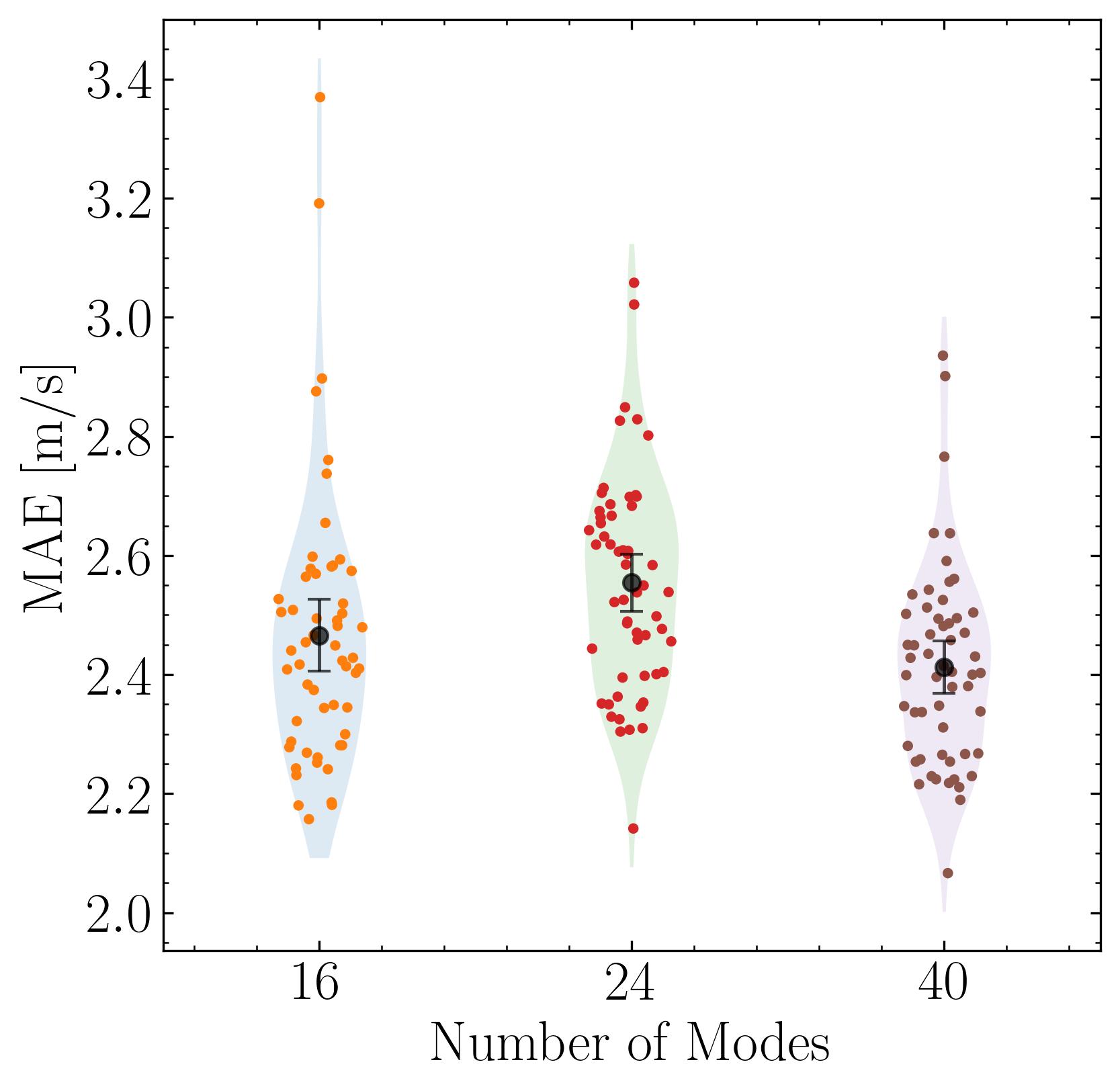}
         \caption{Mean absolute error distribution over the test dataset microstructures ($D_{\mu} = 41 \mu m$; AR $= 1$). Means are shown as black markers with $95\%$ CIs; shaded bands show kernel density estimates. }
         \label{fig_app: FNO_mode_scatter}
\end{figure}
In Fig.~\ref{fig_app: FNO_mode_scatter}, it can be observed that increasing the number of modes from $16$ to $24$ does not result in a significant increase in accuracy as the mean error in the test dataset remains close. However, $16$ modes show a large scatter compared to the other cases. At $40$ modes, the error is significantly reduced compared to $16$ and $24$. But it should be noted that the least MAE in all the cases is still higher than the mean MAE from U-Net and U-FNO, as seen in Fig.~\ref{fig:MAE}. The possible reason for low accuracy could be the requirement of more data or FNO is not suitable for learning such localized features. This analysis could be further refined by using multiple intialization of FNO and averaging the error over them which could demonstrate clear variations.
 \bibliographystyle{elsarticle-num} 
 \bibliography{elsarticle-template-num}





\end{document}